\documentclass[twocolumn,showkeys]{revtex4}
\usepackage[utf8]{inputenc}
\usepackage{graphicx}

\begin{document}

\title{Hydrodynamic and kinetic representation of the microscopic dynamics
as the transitions on the macroscopic scale of description and meaning of the self-consistent field approximation in these models}

\author{Pavel A. Andreev}
\email{andreevpa@physics.msu.ru}
\affiliation{Department of General Physics, Faculty of physics, Lomonosov Moscow State University, Moscow, Russian Federation, 119991.}

\date{\today}

\begin{abstract}
The open problem of derivation of the relativistic Vlasov equation for the systems of charged particles moving with the velocities up to the speed of light
and creating the electromagnetic field in accordance with the full set of the Maxwell equations is considered.
Moreover, the method of derivation is firstly illustrated on the nonrelativistic kinetic model.
Independent derivation of the relativistic hydrodynamics is also demonstrated.
Key role of these derivations of the hydrodynamic and kinetic equations includes the explicit operator of averaging on the
physically infinitesimal volume suggested by L.S. Kuzmenkov.
\end{abstract}


\maketitle





\section{Introduction}

The selfconsistent field approximation or the meanfield approximation is well established in plasma physics
\cite{Weinberg Gr 72}.
The Vlasov equation and corresponding hydrodynamics
(nonrelativistic five moments, nonrelativistic thirteen moments, and relativistic five moments limits)
are considered in literature in order to study collective plasma phenomena developing on the time scale,
where the contribution of "collisions" is negligible.

The derivation of the Vlasov equation
\cite{Vlasov JETP 38}, \cite{Aleksandrov Rukhadze Book}, \cite{Akhiezer}, \cite{Landau Vol X}
based on the atomic structure of matter
(composition of plasmas as the number of individual electrons and ions)
is developed by N.N. Bogoliubov in 1946
(let us make references on representation of this work in textbooks \cite{Akhiezer}, \cite{Landau Vol X}).
This derivation is made for the nonrelativistic motion of charged particles interacting via the Coulomb interaction.
So, the equations of field are reduced to the Poisson equation.
Further generalization of this approach is made in works
\cite{Zaslavskii JAMTP 66}, \cite{Pavlotskii DAN 73}, \cite{Orlov MM 89},
where the weakly relativistic effects are considered.
It includes the account of the interaction of moving charges via the magnetic field created
in accordance with the Biot-Savart law.
Hence, the problem of kinetic description of particles
which create electromagnetic field in accordance with the full Maxwell equations is
the open problem of up to date existing kinetic theory.
Some important steps in the direction of solving of this problem are made
in works of Y.L. Klimontovich and L.S. Kuzmenkov.
Y.L. Klimontovich suggested the microscopic form of the concentration (and other hydrodynamic functions)
and the distribution function,
which are constructed of the Dirac delta functions (see for instance \cite{Klimontovich book}).
Y.L. Klimontovich also found kinetic equation for the microscopic distribution function.
However, complete derivation requires the transition to the macroscopic scale.
Method of the explicit transition on the macroscopic scale
via the integral operator is suggested by L.S. Kuzmenkov in the set of works,
which includes \cite{Drofa TMP 96}, \cite{Andreev PIERS 2012}, \cite{Kuzmenkov CM 15}
(see also an earlier work on kinetics \cite{Kuz'menkov 91}).

There are several definitions of the average value in the classical statistical physics \cite{Klimontovich Plasma}.
They are the time average and the phase average.
There is also the ergodic hypothesis,
which states that these averages give the same result.
The phase average can give different limit regimes
depending on the number of controlled parameters.
If we know the initial conditions of all particles in each system of the ensemble
we get full dynamic description of each system.
In this regime we do not need to use the ensemble of physical systems at all.
We can consider the dynamical evolution of the single system.


This regime, to some extend, resembles the analysis given in this paper.
However, inspite the similarity, there is the fundamental difference.
If we consider any physical system
it does not "know"
which level of knowledge of this system we have.
Our goal is to consider the evolution of this single system.
This evolution is governed by the interaction between particles.
Analysis is made for the structureless particles as the electrons and protons
(we do not need to consider such microscopical particles as quarks on the chosen energy scale)
If there is some structure
which can be modified on the chosen,
like the ionization of atom or ion,
excitation of electron inside of ion, etc,
we have additional dependence on the structure of particles.
Hence, we introduce some microscopic function.
For instance, the concentration
if we want to study the evolution in the coordinate physical space,
or,
the distribution function
if we want to study the evolution in the six-dimensional coordinate -momentum phase space.

The concentration is the deterministic scalar field.
Being the field means that this is the distribution of the physical parameter (characteristic) in the physical space arithmetized by $\textbf{r}$.
The change of this distribution in time corresponds to the motion of particles
via the dependence on coordinates of particles $\textbf{r}_{i}(t)$ on time $t$.
This function describes the exact evolution of the system,
but this evolution is given via the collective variables describing the whole system.
The hydrodynamics itself can be used instead of the mechanical laws of motion for each particle,
but the number of requires hydrodynamic functions
(the concentration, the three projections of the velocity field, etc)
should be equal to the number of degrees of freedom of the system.

The relativistic plasmas are actively studied for the classical and quantum regimes
\cite{Kuz'menkov 91}, \cite{Hakim AoP 82}, \cite{Hakim PRD 92}, \cite{Shatashvili ASS 97},
\cite{Shatashvili PoP 99}, \cite{Hazeltine APJ 2002}, \cite{Mahajan PoP 2002},
\cite{Romatschke IJMPE 10}, \cite{Mahajan PoP 2011}, \cite{Comisso PRL 14},
\cite{Shatashvili PoP 20}, \cite{Hakim book Rel Stat Phys},
\cite{Asenjo PoP 11}, \cite{Melrose BOOK 08},
\cite{Melrose JPA 09}, \cite{Bret PoP 11}, \cite{Ivanov Darwin}, \cite{Asenjo NJP 12},
\cite{Melrose JPA 12}, \cite{Ivanov arxiv big 14}, \cite{Dodin PRA 15 First-principle},
\cite{Mendonca PoP 11}, \cite{Zhu PPCF 12}.
Therefore, the detailed analysis of the derivation of the classic hydrodynamics and kinetics is important
for the better understanding of these physical processes.
Recently, some steps in direction of analysis of the relativistic hydrodynamics are made.
However, the contribution of the multipole moments of the physically infinitesimal volume is ignored in previous papers on this subject
\cite{Andreev 2021 05}, \cite{Andreev 2021 09}, \cite{Andreev 2021 10}
(see also Refs. \cite{Andreev 2112}, \cite{Andreev 2202} for the adaptation of this approach for the degenerate plasmas).
The multipole moments and equations for their evolution are partially described in Ref. \cite{Drofa TMP 96}
for the nonrelativistic regime,
where some nonlinear phenomena are also discussed.



This paper is organized as follows.
In Sec. II the nonrelativistic systems of charged particles are considered on the microscopic scale
and corresponding hydrodynamic equations are obtained in order to make transition on the language
suitable for the description of the collective phenomena.
In Sec. III the selfconsistent field approximation is considered for the nonrelativistic hydrodynamic equations.
In Sec. IV the nonrelativistic kinetic Vlasov equation for the systems of charged particles is derived.
In Sec. V the selfconsistent field approximation is considered for the nonrelativistic kinetic equations.
In Sec. VI The analysis of the selfconsistent field approximation is made for
the relativistic hydrodynamic model with the average reverse gamma factor evolution.
In Sec. VII relativistic kinetic Vlasov equation is derived
and the selfconsistent field approximation is considered for the relativistic kinetic Vlasov equation.
In Sec. VIII a brief summary of obtained results is presented.

\section{Derivation of hydrodynamic equations tracing the microscopic motion of particles}

The self-consistent field or mean-field approximation is well established in the plasmas physics.
However, the derivation of the hydrodynamic and kinetic equations including the explicit operator of averaging on the
physically infinitesimal volume suggested by L.S. Kuzmenkov
\cite{Drofa TMP 96}, \cite{Kuz'menkov 91} allows to give {a deeper look on this approximation.

Our goal in this paper is to consider exact dynamic of the arbitrary system of the charged particles
and represent this dynamical evolution in terms of functions suitable for the collective processes
instead of the parameters characterizes each particle in the system like the coordinates and momentums.

The concentration of particles is traditionally used as one of functions describing the collective effects in the systems of many-particles.
Moreover, the concentration can be introduced as the exact distribution of particles in the coordinate physical space:
\begin{equation}\label{RHD2022ClSCF concentration definition zero volume}
n_{m}(\textbf{r},t)=\sum_{i=1}^{N}\delta(\textbf{r}-\textbf{r}_{i}(t)), \end{equation}
where $\textbf{r}_{i}(t)$ is the coordinate of $i$-th particle,
subindex $m$ in $n_{m}$ refers to the fact
that we consider the microscopic concentration.
We do not know the value of coordinates of particles $\textbf{r}_{i}(t)$.
However, we do not need to know this information.
We need to know equations of motion of each particle.
The equations of motion of particles allow us to derive equations for the evolution of the collective variables.
So, we will discuss the properties of the system in terms of the collective motion with no further references to the coordinates.

We consider the system of classic particles.
We consider the elastic interactions,
so we model the dynamics of all particles as the structureless objects
\begin{equation}\label{RHD2022ClSCF Eq of Motion Newtor nonRel Gen}
\dot{\textbf{p}}_{i}(t)=\textbf{F}(\textbf{r}_{i}(t),t), \end{equation}
where $\textbf{p}_{i}(t)$ is the momentum,
which is the function of time,
and
$\textbf{F}(\textbf{r}_{i}(t),t)$ is the force acting on $i$-th particle being in point $\textbf{r}_{i}(t)$.
All interaction between objects happens via the fields
(usually the electromagnetic field,
while gravitational and nuclear fields are usually give no effects on atomic or plasmas effects).
Therefore, the force is the projection of the corresponding force field on the trajectory of the $i$-th particle
\begin{equation}\label{RHD2022ClSCF projection of FF}
\textbf{F}(\textbf{r}_{i}(t),t)=\int d\textbf{r}\textbf{F}(\textbf{r},t)\delta(\textbf{r}-\textbf{r}_{i}(t)).
\end{equation}
The force $\textbf{F}(\textbf{r}_{i}(t),t)$ is the superposition of interactions with over particles in the system.

Actually we do not need to know the form of interaction to derive equation for evolution of
function (\ref{RHD2022ClSCF concentration definition zero volume}).
Its time derivative gives the following relation
\begin{equation}\label{RHD2022ClSCF cont n-m} \partial_{t}n_{m}+\nabla \cdot\textbf{j}_{m}=0,
\end{equation}
where
\begin{equation}\label{RHD2022ClSCF current definition zero volume}
\textbf{j}_{m}(\textbf{r},t)=\sum_{i=1}^{N}\textbf{v}_{i}(t)\delta(\textbf{r}-\textbf{r}_{i}(t)) \end{equation}
is the microscopic current,
$\textbf{v}_{i}(t)=\dot{\textbf{r}}_{i}(t)$ is the velocity of $i$-th particles,
and we also assume that each particle is stable,
it does not decay on other particles during evolution.
The processes of creation/annoholation or ionization/recombination are not included in our analysis.

Concentration (\ref{RHD2022ClSCF concentration definition zero volume}) is the collective variable.
However, it is constructed on the microscopic scale.
Since, function shows exact position of each particle in some point of space.

We can make the transition to the macroscopic scale.
To this end, we need to introduce the scale giving the macroscopically infinitesimal volume.
We use notation $\Delta$ for this volume.

In each moment of time $t$,
we consider each point of space $\textbf{r}$.
We construct the $\Delta$ vicinity of each point of space $\textbf{r}$
and calculate the concentration on the chosen scale
\begin{equation}\label{RHD2022ClSCF concentration definition via N r t}
n(\textbf{r},t)=\frac{N(\textbf{r},t)}{\Delta}. \end{equation}
However, we do not know the number of particles $N(\textbf{r},t)$ in the $\Delta$ vicinity of any point.
So, this definition is not useful for the further calculation.

\begin{figure}\includegraphics[width=8cm,angle=0]{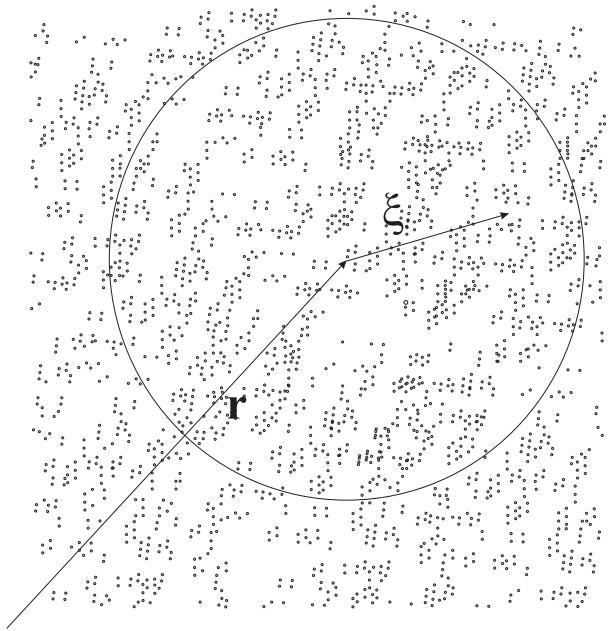}
\caption{\label{RHD2022ClSCF Fig 01}
The delta vicinity is illustrated.
Vector $\mbox{\boldmath$\xi$}$ scanning $\Delta$-vicinity is illustrated.}
\end{figure}

Next, we make the proper generalization of definition (\ref{RHD2022ClSCF concentration definition via N r t})
\cite{Drofa TMP 96}:
\begin{equation}\label{RHD2022ClSCF concentration definition}
n(\textbf{r},t)=\frac{1}{\Delta}\int_{\Delta}d\mbox{\boldmath $\xi$}\sum_{i=1}^{N/2}\delta(\textbf{r}+\mbox{\boldmath $\xi$}-\textbf{r}_{i}(t)). \end{equation}
The $\Delta$-vicinity presented in this formula is illustrated in Fig. \ref{RHD2022ClSCF Fig 01}.

The hydrodynamic model of plasmas requires the introduction of the concentration for each species.
Each concentration and other hydrodynamic functions evolve under the action of all species in the system.
We specify that we consider the quasi-neutral plasmas of two species: the electron-ion plasmas.
We use the following numeration of particles: $i\in [1,N/2]$ for the electrons
and $i\in [N/2+1,N]$ for the ions.
We illustrate the derivation following the evolution of the concentration of electrons.
The explicit contribution of ions is shown in the terms describing the interaction.

Equation (\ref{RHD2022ClSCF concentration definition}) can be interpreted as action of operator of averaging
\cite{Drofa TMP 96}, \cite{Andreev PIERS 2012}:
\begin{equation}\label{RHD2022ClSCF formula for average}\langle ...\rangle\equiv\frac{1}{\Delta}\int_{\Delta}d\mbox{\boldmath $\xi$}
\sum_{i=1}^{N} ... \delta(\textbf{r}+\mbox{\boldmath $\xi$}-\textbf{r}_{i}(t)),\end{equation}
which calculates the number of particles in the $\Delta$-vicinity checking presence of each particle in the chosen vicinity
scanning the vicinity by means vector $\mbox{\boldmath $\xi$}$.
Operator (\ref{RHD2022ClSCF formula for average}) can be replaced by symbol $\langle ...\rangle$ to express equations in shorter form.

This method is suggested by Kuz'menkov L.S.,
it appears as the generalization of method suggested by Klimontovich Yu.L.
\cite{Klimontovich Plasma}, \cite{Klimontovich Dokl 62}, \cite{Weinberg Gr 72}.
However, as it is mentioned in the Introduction,
there is some different physical insite.
Since, we introduce the space average on the physically infinitesimal volume
in contrast with average on the ensemble of physical systems.


From physical point of view, there is the question on the value of the $\Delta$-vicinity.
The problem of estimation of the physically infinitesimal volume is addressed in literature.
For example, Klimontovich Yu.L. in his book \cite{Klimontovich book}
(see also \cite{Klimontovich Plasma}) gives this estimation for two regimes:
the rarefied gas of neutral atoms, and the rarefied plasmas.
This estimation shows that the characteristic length for plasmas is of order of
the Debye length $r_{D}=\sqrt{T/4\pi\sum_{s}n_{0s}q_{s}^{2}}$.
However, this clear statement contains a contradiction for the hydrodynamics.
Nevertheless, this estimation confirms below for the kinetic model.
The physically infinitesimal volume has some nonzero value from microscopic point of view,
but it is the zero volume on the macroscopic scale.
So, it should be expressed via microscopic parameters related to the motion of individual particles.
Let us specify that some authors refer to the notion "microscopic" as to the kinetic description,
while the macroscopic is reserved for the hydrodynamics.
In contrast to it, we call "microscopic" the scale,
where we consider the motion of the individual particles,
while both the kinetic and hydrodynamic models are considered as the macroscopic.
Hence, the macroscopic scale is the scale,
where the concentration (\ref{RHD2022ClSCF concentration definition})
and the distribution function presented below are continuous functions
(as much continuous as they can be in the world of discrete particles and atoms).
The calculation of the Debye length is made on the macroscopic scale.
The expression of the Debye length is found via the macroscopic parameters as well
(so any attempt to find microscopic derivation does not change our conclusion).
Hence, it define some macroscopical volume
which does not corresponds to the macroscopically zero volume of the physically infinitesimal volume.
Nevertheless, it is essential to have large number of particles in the physically infinitesimal volume
to insure the continuity of hydrodynamic functions.
For the low density plasmas we expect to have the following
relations of the characteristic lengths:
$r_{B}\ll a\equiv (\sqrt[3]{n})^{-1}\ll \sqrt[3]{\Delta}\ll r_{D}$,
where $r_{B}=\hbar^{2}/m_{e}e^{2}$ is the Bohr radius.
For the high density plasmas we have modified
relations of the characteristic lengths:
$r_{B}\sim a\ll \sqrt[3]{\Delta}\ll r_{D}$.

Similarly to equations (\ref{RHD2022ClSCF concentration definition zero volume}),
(\ref{RHD2022ClSCF cont n-m}), and (\ref{RHD2022ClSCF current definition zero volume})
we can derive the continuity equation for the concentration (\ref{RHD2022ClSCF concentration definition}).
In order to get the derivation,
we differentiate expression (\ref{RHD2022ClSCF concentration definition}) with respect to time
and obtain the continuity equation
\begin{equation}\label{RHD2022ClSCF cont via j}
\partial_{t}n+\nabla\cdot\textbf{j}=0,\end{equation}
where the current $\textbf{j}$ has the following definition
\begin{equation}\label{RHD2022ClSCF current definition}
\textbf{j}(\textbf{r},t)=
\frac{1}{\Delta}\int_{\Delta}d\mbox{\boldmath $\xi$}\sum_{i=1}^{N/2} \textbf{v}_{i}(t)
\delta(\textbf{r}+\mbox{\boldmath $\xi$}-\textbf{r}_{i}(t)). \end{equation}
We can also introduce the velocity field
$\textbf{v}=\textbf{j}/n$.
For the current of the single species,
we find that the momentum density $\textbf{P}$ is proportional to the current $\textbf{P}=m \textbf{j}$.

In order to continue the derivation of hydrodynamics
we consider the time evolution of the current $\textbf{j}$.
In this case we need the expression for the acceleration
with the explicit form of interaction
$$\dot{\textbf{v}}_{i}(t)=\frac{1}{m_{i}}
\biggl(q_{i}\textbf{E}_{ext}(\textbf{r}_{i}(t),t)$$
\begin{equation}\label{RHD2022ClSCF Eq of Motion Newton nonRel}
+\frac{1}{c}q_{i}[\textbf{v}_{i}(t), \textbf{B}_{ext}(\textbf{r}_{i}(t),t)]
-\sum_{j=1, j\neq i}^{N}q_{i}q_{j}\nabla_{i}G_{ij}\biggr), \end{equation}
where $G_{ij}=1/r_{ij}$ is the Green function of the Coulomb interaction,
$r_{ij}=\mid \textbf{r}_{i}(t)-\textbf{r}_{j}(t)\mid$.
The average velocity is found as arithmetic mean for all particles being in the vicinity.

In this section we consider the nonrelativistic plasmas.
Therefore, we consider interaction in the quasi-static limit,
which is presented by the Coulomb interaction.

The action of the time derivative on the current (\ref{RHD2022ClSCF current definition})
leads to the action of the time derivative on the product of two functions under the integral
$\textbf{v}_{i}(t)
\delta(\textbf{r}+\mbox{\boldmath $\xi$}-\textbf{r}_{i}(t))$.
Hence, the result is the superposition of two terms under the integral
$\dot{\textbf{v}}_{i}(t)
\delta(\textbf{r}+\mbox{\boldmath $\xi$}-\textbf{r}_{i}(t))$
requiring the acceleration (\ref{RHD2022ClSCF Eq of Motion Newton nonRel}),
and
$-\textbf{v}_{i}(t)
(\textbf{v}_{i}(t)\cdot\nabla)
\delta(\textbf{r}+\mbox{\boldmath $\xi$}-\textbf{r}_{i}(t))$
leading to the flux of the momentum.

It gives the general structure of the Euler equation or the momentum balance evolution equation
\begin{equation}\label{RHD2022ClSCF Euler equation for j non rel General}
\partial_{t}j^{a}+\partial_{b}\Pi^{ab}=\frac{1}{m}(\Phi^{a}_{ext}+\Phi^{a}) ,\end{equation}
where
\begin{equation}\label{RHD2022ClSCF} \Pi^{ab}(\textbf{r},t)=
\frac{1}{\Delta}\int_{\Delta}d\mbox{\boldmath $\xi$}\sum_{i=1}^{N/2} v_{i}^{a}(t) v_{i}^{b}(t)
\delta(\textbf{r}+\mbox{\boldmath $\xi$}-\textbf{r}_{i}(t)) \end{equation}
is the momentum flux,
$$\mbox{\boldmath $\Phi$}_{ext}=\frac{1}{\Delta}\int_{\Delta}d\mbox{\boldmath $\xi$}\sum_{i=1}^{N/2}
\biggl(q_{i}\textbf{E}_{ext}(\textbf{r}_{i}(t),t)$$
\begin{equation}\label{RHD2022ClSCF Phi ext def} +\frac{1}{c}q_{i}[\textbf{v}_{i}(t), \textbf{B}_{ext}(\textbf{r}_{i}(t),t)]\biggr)
\delta(\textbf{r}+\mbox{\boldmath $\xi$}-\textbf{r}_{i}(t)) \end{equation}
is the density of the force caused by the action of the external fields,
\begin{equation}\label{RHD2022ClSCF Phi inter def} \mbox{\boldmath $\Phi$}=-\frac{1}{\Delta}\int_{\Delta}d\mbox{\boldmath $\xi$}\sum_{i=1}^{N/2}
\sum_{j=1,j\neq i}^{N}q_{i}q_{j}\nabla_{i}G_{ij}
\delta(\textbf{r}+\mbox{\boldmath $\xi$}-\textbf{r}_{i}(t)) \end{equation}
is the density of the force caused by the interparticle interaction,
$a$ and other Latin indexes (from beginning of alphabet) correspond to vector indexes in the Euclidian space,
the Einstein rule of the summation on the repeating indexes is assumed:
$\textbf{j} \cdot\textbf{E}=\sum_{a} j_{a} E^{a}=j_{a} E^{a}=j^{a} E^{a}$.

\subsection{Multipole moments of physically infinitesimal volume in the external force field}

We consider the external force field as two parts,
one related to the electric filed
\begin{equation}\label{RHD2022ClSCF electric part FF i}
\mbox{\boldmath $\Phi$}_{ext,el}=q_{s}\frac{1}{\Delta}\int_{\Delta}d\mbox{\boldmath $\xi$}\sum_{i=1}^{N/2}
\textbf{E}_{ext}(\textbf{r}_{i}(t),t) \delta_{i}, \end{equation}
and the second related to the magnetic field
\begin{equation}\label{RHD2022ClSCF magnetic part FF i} \mbox{\boldmath $\Phi$}_{ext,m}
=\frac{q_{s}}{c}\frac{1}{\Delta}\int_{\Delta}d\mbox{\boldmath $\xi$}\sum_{i=1}^{N/2}
[\textbf{v}_{i}(t), \textbf{B}_{ext}(\textbf{r}_{i}(t),t)] \delta_{i}, \end{equation}
where
$\delta_{i}\equiv \delta(\textbf{r}+\mbox{\boldmath $\xi$}-\textbf{r}_{i}(t))$ is the short notation.

Start our discussion with the electric part of the external force field.
We use the delta-function under the integral to represent the argument of the electric field
\begin{equation}\label{RHD2022ClSCF electric part FF r xi}
\mbox{\boldmath $\Phi$}_{ext,el}=q_{s}\frac{1}{\Delta}\int_{\Delta}d\mbox{\boldmath $\xi$}\sum_{i=1}^{N/2}
\textbf{E}_{ext}(\textbf{r}+\mbox{\boldmath $\xi$},t) \delta_{i}, \end{equation}
where we cannot place the electric field outside of the integral.
However, if the electric field changes slowly over the physically infinitesimal volume
(over the $\Delta$-vicinity)
we can expand function $\textbf{E}_{ext}(\textbf{r}+\mbox{\boldmath $\xi$},t)$ on the vector $\mbox{\boldmath $\xi$}$ scanning the $\Delta$-vicinity.
Keeping few major terms of the expansion
we find
$$\textbf{E}_{ext}(\textbf{r}+\mbox{\boldmath $\xi$},t)
\approx
\textbf{E}_{ext}(\textbf{r},t)$$
\begin{equation}\label{RHD2022ClSCF expansion of E}
+(\mbox{\boldmath $\xi$}\cdot\nabla) \textbf{E}_{ext}(\textbf{r},t)
+\frac{1}{2}(\mbox{\boldmath $\xi$}\cdot\nabla)^{2} \textbf{E}_{ext}(\textbf{r},t)+...\end{equation}
The substitution of this expression in the electric part of the external force field
(\ref{RHD2022ClSCF electric part FF r xi})
gives corresponding expression of the electric part of the external force field
$$\mbox{\boldmath $\Phi$}_{ext,el}=q_{s} n_{s}\textbf{E}_{ext}(\textbf{r},t)$$
\begin{equation}\label{RHD2022ClSCF} +q_{s} (\textbf{d}\cdot\nabla)\textbf{E}_{ext}(\textbf{r},t)
+q_{s}Q^{ab}\partial_{a}\partial_{b}\textbf{E}_{ext}(\textbf{r},t)
+...\end{equation}
where
\begin{equation}\label{RHD2022ClSCF}
\textbf{d}\equiv \textbf{d}(\textbf{r},t)
=\frac{1}{\Delta}\int_{\Delta}d\mbox{\boldmath $\xi$}\sum_{i=1}^{N/2}
\mbox{\boldmath $\xi$}\delta(\textbf{r}+\mbox{\boldmath $\xi$}-\textbf{r}_{i}(t)) \end{equation}
is the electric dipole moment of the $\Delta$-vicinity devided by the charge $q_{s}$,
and
\begin{equation}\label{RHD2022ClSCF} Q^{ab}(\textbf{r},t)=\frac{1}{\Delta}\int_{\Delta}d\mbox{\boldmath $\xi$}\sum_{i=1}^{N/2}
\xi^{a}\xi^{b}\delta(\textbf{r}+\mbox{\boldmath $\xi$}-\textbf{r}_{i}(t)) \end{equation}
is the electric quadrupole moment of the $\Delta$-vicinity devided by the charge $q_{s}$.


\subsection{Multipole moments of the Lorentz force field}

Let us consider the magnetic part of the force field (\ref{RHD2022ClSCF magnetic part FF i}),
where we expand the magnetic field on the vector $\mbox{\boldmath $\xi$}$ scanning the $\Delta$-vicinity:
$$\textbf{B}_{ext}(\textbf{r}_{i}(t),t)=\textbf{B}_{ext}(\textbf{r}+\mbox{\boldmath $\xi$},t)
\approx
\textbf{B}_{ext}(\textbf{r},t)$$
\begin{equation}\label{RHD2022ClSCF expansion of B}
+(\mbox{\boldmath $\xi$}\cdot\nabla) \textbf{B}_{ext}(\textbf{r},t)
+\frac{1}{2}(\mbox{\boldmath $\xi$}\cdot\nabla)^{2} \textbf{B}_{ext}(\textbf{r},t)+...\end{equation}

Therefore, equation (\ref{RHD2022ClSCF magnetic part FF i}) can be rewritten as
$$\Phi_{ext,m}^{a}
=\frac{q_{s}}{c}\varepsilon^{abc}\biggl(j^{b} B_{ext}^{c}(\textbf{r},t)$$
\begin{equation}\label{RHD2022ClSCF magnetic part FF i}
+ J_{D}^{bd}\partial^{d}B_{ext}^{c}(\textbf{r},t)
+J_{Q}^{bdf} \partial^{d}\partial^{f}B_{ext}^{c}(\textbf{r},t)\biggr)+..., \end{equation}
where we also use the Levi-Civita symbol $\varepsilon^{abc}$ for the vector product in the tensor notations
\begin{equation}\label{RHD2022ClSCF}
J_{D}^{ab}(\textbf{r},t)
=\frac{1}{\Delta}\int_{\Delta}d\mbox{\boldmath $\xi$}\sum_{i=1}^{N/2}
v_{i}^{a}(t)\xi^{b}\delta(\textbf{r}+\mbox{\boldmath $\xi$}-\textbf{r}_{i}(t)) \end{equation}
is the flux of the electric dipole moment of the $\Delta$-vicinity devided by the charge $q_{s}$,
and
\begin{equation}\label{RHD2022ClSCF}
J_{Q}^{abc}(\textbf{r},t)=\frac{1}{\Delta}\int_{\Delta}d\mbox{\boldmath $\xi$}\sum_{i=1}^{N/2}
v_{i}^{a}(t)\xi^{b}\xi^{c}\delta(\textbf{r}+\mbox{\boldmath $\xi$}-\textbf{r}_{i}(t)) \end{equation}
is the flux of the electric quadrupole moment of the $\Delta$-vicinity devided by the charge $q_{s}$.

\section{The selfconsistent field approximation in non-relativistic hydrodynamics}

In the previous section we presented derivation of the general form of the Euler equation.
Moreover, we considered the multipole expansion of the external force field.
Here, we consider the interparticle interaction.
We have two goals too achive in this section.
The first goal is the analysis of the selfconsistent field approximation
in order to understand its properties in the deterministic derivation of the hydrodynamic equations.
The second goal is the multipole expansion of the interparticle interaction force field.


We repeat equation (\ref{RHD2022ClSCF Phi inter def}) with underlying some functions under the integral.
Moreover, it is useful to specify number of species in the system.
To get most simple presentation
we chose the electron-proton plasmas
(or completely ionized hydrogen plasmas).
Let us numerate electrons as the particles with numbers $i\in [1,N/2]$
and ions as the particles with numbers $i\in [N/2+1,N]$.
Let us also to point out that
the set of hydrodynamic equations is obtained for each species.
We focus on dynamics of electrons.
Therefore, the force field acting on electrons is composed of the electron-electron interaction
and the force field created by ions and acting on the electrons.
The force field created by electrons,
which acts on the electrons (the selfaction of the electron material field),
has the following form
\begin{equation}\label{RHD2022ClSCF Phi inter def represented ee}
\mbox{\boldmath $\Phi$}_{e-e}=-\frac{q_{e}^{2}}{\Delta}\int_{\Delta}d\mbox{\boldmath $\xi$}\sum_{i,j=1,j\neq i}^{N/2}
\nabla_{i}G (\mid \textbf{r}_{i}(t)-\textbf{r}_{j}(t)\mid)
\delta(\textbf{r}+\mbox{\boldmath $\xi$}-\textbf{r}_{i}(t)). \end{equation}
The action of ions on the electrons can be written in the following form
$$\mbox{\boldmath $\Phi$}_{e-i}=-q_{e}q_{i}\frac{1}{\Delta}\int_{\Delta}d\mbox{\boldmath $\xi$}\times$$
\begin{equation}\label{RHD2022ClSCF Phi inter def represented ei}
\times\sum_{i=1}^{N/2}\sum_{j=N/2+1}^{N}
\nabla_{i}G (\mid \textbf{r}_{i}(t)-\textbf{r}_{j}(t)\mid)
\delta(\textbf{r}+\mbox{\boldmath $\xi$}-\textbf{r}_{i}(t)). \end{equation}

We continue our analysis for the expression (\ref{RHD2022ClSCF Phi inter def represented ee}).
This force field is not symmetric relatively $i$-th and $j$-th particles.
Introducing the integral over whole scape we include the delta function containing coordinate of $j$-th particle
$$\mbox{\boldmath $\Phi$}_{e-e}=-\frac{q_{e}^{2}}{\Delta}\int d\textbf{r}'\int_{\Delta}d\mbox{\boldmath $\xi$}\sum_{i,j=1,j\neq i}^{N/2}
\nabla_{i}G (\mid \textbf{r}_{i}(t)-\textbf{r}'\mid)\times$$
\begin{equation}\label{RHD2022ClSCF Phi inter def represented ee int II}
\times\delta(\textbf{r}+\mbox{\boldmath $\xi$}-\textbf{r}_{i}(t))
\delta(\textbf{r}'-\textbf{r}_{j}(t)). \end{equation}
Here we see that the delta functions containing the $i$-th and $j$-th particles have different structure of arguments.
We need to continue the symmetrization of the force field.
To this end, we use the following mathematical relation:
if we have two functions with the following relation
$f(\textbf{r})=(1/\Delta) \int_{\Delta}g(\textbf{r}+\mbox{\boldmath $\xi$}) d\mbox{\boldmath $\xi$}$
we find that their integrals over the whole space are equal to each other
$\int d\textbf{r} f(\textbf{r})=\int d\textbf{r} g(\textbf{r})$.
Consequently, we can represent integral
$\int d\textbf{r}'\nabla_{i}G (\mid \textbf{r}_{i}(t)-\textbf{r}'\mid) \delta(\textbf{r}'-\textbf{r}_{j}(t))$
as the following structure
$(1/\Delta)\int d\textbf{r}' \int_{\Delta}d\mbox{\boldmath $\xi$}
\nabla_{i}G (\mid \textbf{r}_{i}(t)-\textbf{r}'-\mbox{\boldmath $\xi$}\mid) \delta(\textbf{r}'+\mbox{\boldmath $\xi$}-\textbf{r}_{j}(t))$.
It leads to the symmetric form of
$$\mbox{\boldmath $\Phi$}_{e-e}=-\frac{q_{e}^{2}}{\Delta^{2}}
\int d\textbf{r}'\int_{\Delta}d\mbox{\boldmath $\xi$}\int_{\Delta}d\mbox{\boldmath $\xi$}'
\sum_{i,j=1,j\neq i}^{N/2}\times$$
\begin{equation}\label{RHD2022ClSCF Phi inter def represented ee int III}
\times\nabla_{i}G (\mid \textbf{r}_{i}(t)-\textbf{r}_{j}(t)\mid)
\delta(\textbf{r}+\mbox{\boldmath $\xi$}-\textbf{r}_{i}(t))
\delta(\textbf{r}'+\mbox{\boldmath $\xi$}'-\textbf{r}_{j}(t)). \end{equation}

\begin{figure}\includegraphics[width=8cm,angle=0]{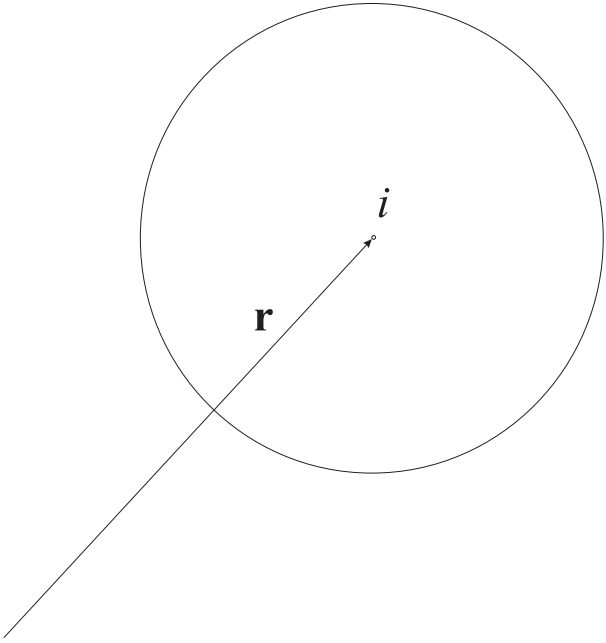}
\caption{\label{RHD2022ClSCF Fig 02}
The $\Delta$-vicinity around arbitrary $i$-th particle is illustrated.} \end{figure}

\begin{figure}\includegraphics[width=8cm,angle=0]{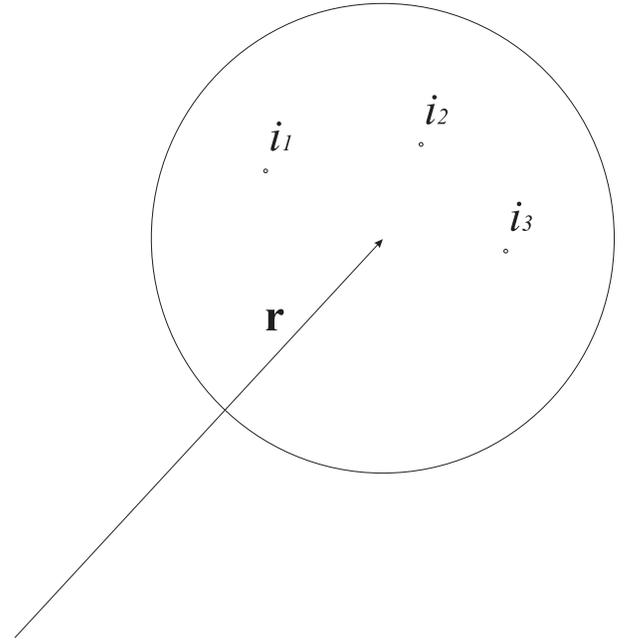}
\caption{\label{RHD2022ClSCF Fig 03}
The $\Delta$-vicinity around arbitrary point of space $\textbf{r}$ is pictured.
While some particles are around the center.} \end{figure}

\subsection{Monopole approximation and the selfconsistent field approximation}

We can use the delta function to express coordinates of particles in the Green function on
$\textbf{r}+\mbox{\boldmath $\xi$}$ and $\textbf{r}'+\mbox{\boldmath $\xi$}'$.
$$\mbox{\boldmath $\Phi$}_{e-e}=-\frac{q_{e}^{2}}{\Delta^{2}}
\int d\textbf{r}'\int_{\Delta}d\mbox{\boldmath $\xi$}\int_{\Delta}d\mbox{\boldmath $\xi$}'
\sum_{i,j=1,j\neq i}^{N/2}
\times$$
\begin{equation}\label{RHD2022ClSCF Phi inter def represented ee int IV}
\times\nabla_{i}G (\mid \textbf{r}+\mbox{\boldmath $\xi$}-\textbf{r}'-\mbox{\boldmath $\xi$}'\mid)
\delta(\textbf{r}+\mbox{\boldmath $\xi$}-\textbf{r}_{i}(t))
\delta(\textbf{r}'+\mbox{\boldmath $\xi$}'-\textbf{r}_{j}(t)). \end{equation}
Both expressions (\ref{RHD2022ClSCF Phi inter def represented ee int III}) and (\ref{RHD2022ClSCF Phi inter def represented ee int IV})
show
that we cannot introduce the two-particle concentration on this stage of the derivation.

First, we consider the multipole expansion of the Green function
$G (\mid \textbf{r}+\mbox{\boldmath $\xi$}-\textbf{r}'-\mbox{\boldmath $\xi$}'\mid)$
assuming that
it slowly changes on the scale of the $\Delta$-vicinity.
For simplicity, in this subsection, we consider the zero order expansion (the monopole limit).
So, we have
$G (\mid \textbf{r}+\mbox{\boldmath $\xi$}-\textbf{r}'-\mbox{\boldmath $\xi$}\mid)
\approx G (\mid \textbf{r}-\textbf{r}'\mid)$.
Hence, the Green function can be placed out the integral on the $\Delta$-vicinities:
\begin{equation}\label{RHD2022ClSCF Phi inter via n2}
\mbox{\boldmath $\Phi$}_{e-e}=-q_{e}^{2}
\int d\textbf{r}'
\nabla_{\textbf{r}}G (\mid \textbf{r}-\textbf{r}'\mid)
\cdot n_{2,ee}(\textbf{r},\textbf{r}',t), \end{equation}
where
$$n_{2,ee}(\textbf{r},\textbf{r}',t)=
\frac{1}{\Delta^{2}}\times$$
\begin{equation}\label{RHD2022ClSCF n 2 def}
\times\int_{\Delta}d\mbox{\boldmath $\xi$}\int_{\Delta}d\mbox{\boldmath $\xi$}'
\sum_{i,j=1,j\neq i}^{N/2}
\delta(\textbf{r}+\mbox{\boldmath $\xi$}-\textbf{r}_{i}(t))
\delta(\textbf{r}'+\mbox{\boldmath $\xi$}'-\textbf{r}_{j}(t))
\end{equation}
is the two-particle concentration.

The Debay radius $r_{D}$ is the distance,
where the Coulomb field of the charge is screened.
The screening is a macroscopic effect
which requires the macroscopic number of particles in the Debay sphere.
The average interparticle distance is $a\equiv n^{-1/3}$
and we have $r_{D}\gg a$.

If we consider the neutral particles
we have strong decrease of the potential of interaction.
Hence, the considerable changes in the state of motion of neutral particles are interpreted as the collisions
since it happens at the small aiming parameter.
The neutral particles move as the free particles between collisions.

In plasmas we have the long-range interaction.
The interaction is screened,
but it happens on the scale of the Debay radius $r_{D}$.
However, the interaction of the charged particles being inside the Debay sphere is not interpreted as the collisions,
at least not all of these interactions have interpretation as these collisions,
but the small part of them.

There is the mechanism of chaotic interactions
which transits the system to the equilibrium state
and leads to the increase of the entropy.
It corresponds to the interaction at the small interparticle distances.
It can be interpreted as the scattering with the small aiming parameter.
While the interaction on the large distances do not lead to the relaxation (see also \cite{Landau Vol X}).

\begin{figure}\includegraphics[width=8cm,angle=0]{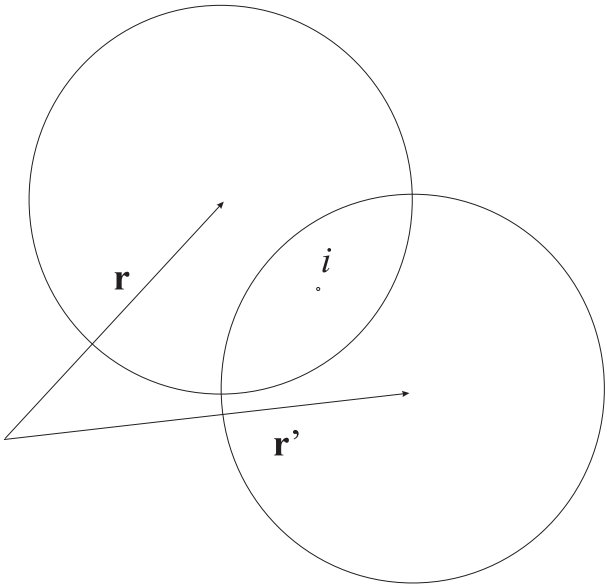}
\caption{\label{RHD2022ClSCF Fig 04}
The interaction is represented via the two-particle functions,
which includes the consideration of delta vicinities of two arbitrary points $\textbf{r}$ and $\textbf{r}'$.
Figure illustrates the regime of overlapping delta vicinities of points $\textbf{r}$ and $\textbf{r}'$.
Position of $i$-th particles belonging to both vicinities is illustrated as well.}
\end{figure}


Particles mostly are located at the distances corresponding to the average interparticle distances from their neighbors.
While, the particles move in the average collective field.
However, incident convergence to distances $\Delta r\ll a$ gives strong scattering.
Described scales allow to introduce corresponding scaling of the electromagnetic field
acting on $i$-th particle located at point $\textbf{r}_{i}(t')=\textbf{r}'$.
For instance, let us present the decomposition of the electromagnetic field vector
$\textbf{E}_{i}(\textbf{r}_{i}(t'),t')=\textbf{E}_{>\Delta}(\textbf{r}',t')+\textbf{e}_{<\Delta}(\textbf{r}',t')$.
Here, vector $\textbf{E}_{>\Delta}$ is the electric field created by the particles being beyond the $\Delta$-vicinity surrounding the $i$-th particle
(see Fig. \ref{RHD2022ClSCF Fig 02}).
Vector $\textbf{e}_{<\Delta}$ is the electric field created by the particles being inside the $\Delta$-vicinity surrounding the $i$-th particle.

However, if we consider the $\Delta$-vicinity around the arbitrary point of space $\textbf{r}$
and consider the evolution of particles inside (getting in or out) the $\Delta$-vicinity
we have distribution of particles in different points of the vicinity (not in its center),
like it is demonstrated in Fig. \ref{RHD2022ClSCF Fig 03}.
So, we have the following picture for the arbitrary particle in the $\Delta$-vicinity.

As it is mentioned above ratio $a/r_{D}$ is the small parameter.
However, we want to specify
that it corresponds to the square of the small dimensionless parameter $a/r_{D}=\epsilon^{2}$
in order to introduce the intermediate scale $\Delta^{1/3}$.
It leads to the following explicit expression for the radius of the $\Delta$-vicinity:
$\Delta^{1/3}=\sqrt{a\cdot r_{D}}$.

Fig. \ref{RHD2022ClSCF Fig 04} shows $i$-th picture
which belongs to the $\Delta_{\textbf{r}}$-vicinity of point $\textbf{r}$
and to the $\Delta_{\textbf{r}'}$-vicinity of point $\textbf{r}'$.
Hence, the particle $i$ is under action of "collisions" from the particle $j\in \Delta_{\textbf{r}'}$
(but being beyond $\Delta_{\textbf{r}}$,
see Fig. \ref{RHD2022ClSCF Fig 05}).
Fig. \ref{RHD2022ClSCF Fig 05} shows particles $j_{k}\in\Delta_{\textbf{r}'}$,
but they do not belong to $\Delta_{\textbf{r}}$,
which "collide" with particle $i$ in order to change the momentum of particles in the vicinity $\Delta_{\textbf{r}}$.
Fig. \ref{RHD2022ClSCF Fig 06} shows the $i$-th nd $j$-th particles
which simultaneously belong to $\Delta_{\textbf{r}}$ and $\Delta_{\textbf{r}'}$.
Their interaction do not change the momentum of all particles in the $\Delta_{\textbf{r}}$ due to the Newton's third law.
In order to neglect the contribution of the collisions of
$i$-th and $j$-th particles illustrated in Fig. \ref{RHD2022ClSCF Fig 06} in the evolution of the particles in the $\Delta_{\textbf{r}}$-vicinity
we need to keep $\textbf{r}'$ at distances larger then $2\sqrt[3]{\Delta}$ from point $\textbf{r}$.
So, the vicinities do not cross each other.
Particles $i_{k}\in\Delta_{\textbf{r}}$ interact with particles
$j_{k}\in\Delta_{\textbf{r}'}$ up to distances $\mid \textbf{r}-\textbf{r}'\mid\sim r_{De}$.
The interaction can be neglected completely for the larger distances.

\begin{figure}\includegraphics[width=8cm,angle=0]{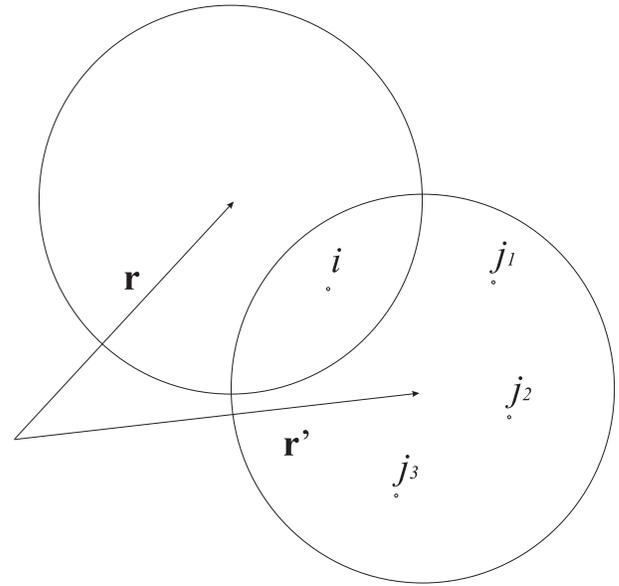}
\caption{\label{RHD2022ClSCF Fig 05}
The two groups of interacting particles are illustrated.
One group is illustrated via the single particle $i$ belonging to both vicinities.
The second group of particles is illustrated by $j_{1}$, $j_{2}$ and $j_{3}$,
which belong to the delta vicinity of point $\textbf{r}'$.}
\end{figure}

\begin{figure}\includegraphics[width=8cm,angle=0]{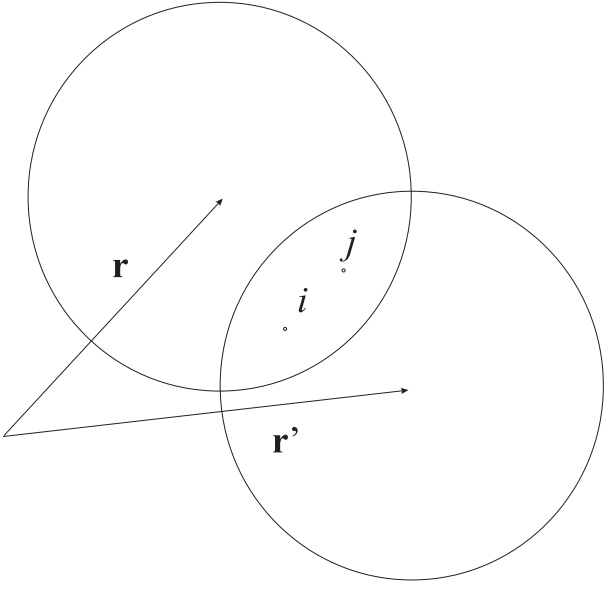}
\caption{\label{RHD2022ClSCF Fig 06}
The two interacting particles simultaneously being parts of two delta-vicinities are illustrated.}
\end{figure}

We use the selfconsistent field approximation in equation (\ref{RHD2022ClSCF Phi inter via n2}),
so we assume
$n_{2}(\textbf{r},\textbf{r}',t)=n(\textbf{r},t)n(\textbf{r}',t)$,
and find
\begin{equation}\label{RHD2022ClSCF Phi inter via n n}
\mbox{\boldmath $\Phi$}_{e-e}=-q_{e}^{2}n(\textbf{r},t)
\nabla_{\textbf{r}}\int d\textbf{r}'
G (\mid \textbf{r}-\textbf{r}'\mid)
n(\textbf{r}',t). \end{equation}
Expression (\ref{RHD2022ClSCF Phi inter via n n}) allows us to introduce the electrostatic potential of the electric field
created by electrons
as $\varphi_{e}(\textbf{r},t)=q_{e}\int d\textbf{r}'
G (\mid \textbf{r}-\textbf{r}'\mid)
n(\textbf{r}',t)$, and corresponding electric field $\textbf{E}_{e}=-\nabla_{\textbf{r}}\varphi_{e}(\textbf{r},t)$.
Obtained electric field satisfies the following equations
$\nabla \times\textbf{E}_{e}=0$, and
$\nabla \cdot\textbf{E}_{e}=4\pi q_{e}n_{e}$.
The complete electric field is the superposition of the electric fields created by all species of the system $\textbf{E}=\sum_{s=e,i}\textbf{E}_{s}$
which obeys the electrostatic Maxwell equations
$\nabla \times\textbf{E}=0$, and
\begin{equation}\label{RHD2022ClSCF div E}  \nabla \cdot\textbf{E}=4\pi \sum_{s}^{} q_{s}n_{s}. \end{equation}

On this stage we can present the intermediate form of the Euler equation (\ref{RHD2022ClSCF Euler equation for j non rel General})
\begin{equation}\label{RHD2022ClSCF Euler equation for j non rel Mon SCF}
\partial_{t}j_{s}^{a}+\partial_{b}\Pi_{s}^{ab}=
\frac{q_{s}}{m_{s}}\biggl(n_{s}(E_{ext}^{a}+E^{a})+\varepsilon^{abc}j_{s}^{b}B_{ext}^{c}\biggr).
\end{equation}

\subsection{Multipole expansion}

In order to consider the multipole expansion
existing
if  the Green function
$G (\mid \textbf{r}+\mbox{\boldmath $\xi$}-\textbf{r}'-\mbox{\boldmath $\xi$}'\mid)$
slowly changes on the scale of the $\Delta$-vicinity,
we need to consider equation (\ref{RHD2022ClSCF Phi inter def represented ee int IV})
in more details.
First, we present the expansion of the Green function
$$G (\mid \textbf{r}-\textbf{r}'+\mbox{\boldmath $\xi$}-\mbox{\boldmath $\xi$}'\mid)
\approx G (\mid \textbf{r}-\textbf{r}'\mid)
+ (\xi^{a}-\xi'^{a})\partial_{a} G (\mid \textbf{r}-\textbf{r}'\mid)$$
\begin{equation}\label{RHD2022ClSCF Green f expansion}
+ \frac{1}{2}(\xi^{a}-\xi'^{a})(\xi^{b}-\xi'^{b}) \partial_{a} \partial_{b} G (\mid \textbf{r}-\textbf{r}'\mid)
+... \end{equation}

The direct substitution gives us the following huge expression
\begin{widetext}
$$\Phi^{a}_{e-e}=-\frac{q_{e}^{2}}{\Delta^{2}}
\int d\textbf{r}'
\partial_{\textbf{r}}^{a}G (\mid \textbf{r}-\textbf{r}'\mid)
\int_{\Delta}d\mbox{\boldmath $\xi$}\int_{\Delta}d\mbox{\boldmath $\xi$}'
\sum_{i,j=1,j\neq i}^{N/2}
\delta(\textbf{r}+\mbox{\boldmath $\xi$}-\textbf{r}_{i}(t))
\delta(\textbf{r}'+\mbox{\boldmath $\xi$}'-\textbf{r}_{j}(t))$$
$$-\frac{q_{e}^{2}}{\Delta^{2}}
\int d\textbf{r}'
\partial_{\textbf{r}}^{a}\partial_{\textbf{r}}^{b}G (\mid \textbf{r}-\textbf{r}'\mid)
\int_{\Delta}d\mbox{\boldmath $\xi$}\int_{\Delta}d\mbox{\boldmath $\xi$}'
\sum_{i,j=1,j\neq i}^{N/2}
(\xi^{a}-\xi'^{a})
\delta(\textbf{r}+\mbox{\boldmath $\xi$}-\textbf{r}_{i}(t))
\delta(\textbf{r}'+\mbox{\boldmath $\xi$}'-\textbf{r}_{j}(t))$$
\begin{equation}\label{RHD2022ClSCF Phi via n2 d2 Q2 via delta}
-\frac{q_{e}^{2}}{\Delta^{2}}
\int d\textbf{r}'
\partial_{\textbf{r}}^{a}\partial_{\textbf{r}}^{b}\partial_{\textbf{r}}^{c}G (\mid \textbf{r}-\textbf{r}'\mid)
\int_{\Delta}d\mbox{\boldmath $\xi$}\int_{\Delta}d\mbox{\boldmath $\xi$}'
\sum_{i,j=1,j\neq i}^{N/2}
\frac{1}{2}(\xi^{a}\xi^{b}-\xi'^{a}\xi^{b}-\xi^{a}\xi'^{b}+\xi'^{a}\xi'^{b})
\delta(\textbf{r}+\mbox{\boldmath $\xi$}-\textbf{r}_{i}(t))
\delta(\textbf{r}'+\mbox{\boldmath $\xi$}'-\textbf{r}_{j}(t))
+... \end{equation}

Obtained expression can be rewritten via corresponding two-particle functions
$$\Phi^{a}_{e-e}=-q_{e}^{2}
\int d\textbf{r}'
\partial_{\textbf{r}}^{a}G (\mid \textbf{r}-\textbf{r}'\mid)
\cdot n_{2}(\textbf{r},\textbf{r}',t)
-q_{e}^{2}
\int d\textbf{r}'
\partial_{\textbf{r}}^{a}\partial_{\textbf{r}}^{b}G (\mid \textbf{r}-\textbf{r}'\mid)
\cdot \biggl(d_{2}^{b}(\textbf{r},\textbf{r}',t)-d_{2}^{b}(\textbf{r}',\textbf{r},t)\biggr)$$
\begin{equation}\label{RHD2022ClSCF Phi via n2 d2 Q2}-\frac{1}{2}q_{e}^{2}
\int d\textbf{r}'
\partial_{\textbf{r}}^{a}\partial_{\textbf{r}}^{b}\partial_{\textbf{r}}^{c}G (\mid \textbf{r}-\textbf{r}'\mid)
\cdot\biggl( Q_{2}^{bc}(\textbf{r},\textbf{r}',t)
+Q_{2}^{bc}(\textbf{r}',\textbf{r},t)
-D_{2}^{bc}(\textbf{r},\textbf{r}',t)-D_{2}^{cb}(\textbf{r},\textbf{r}',t)\biggr), \end{equation}
where
we introduce three two-particles functions
\begin{equation}\label{RHD2022ClSCF d 2 def}
d_{2}^{b}(\textbf{r},\textbf{r}',t)=
\frac{1}{\Delta^{2}}\int_{\Delta}d\mbox{\boldmath $\xi$}d\mbox{\boldmath $\xi$}'
\sum_{i=1,j\neq i}^{N/2}\xi^{b}
\delta(\textbf{r}+\mbox{\boldmath $\xi$}-\textbf{r}_{i}(t))
\delta(\textbf{r}'+\mbox{\boldmath $\xi$}'-\textbf{r}_{j}(t)),
\end{equation}
the permutation of its arguments leads to
\begin{equation}\label{RHD2022ClSCF d 2 def permuted}
d_{2}^{b}(\textbf{r}',\textbf{r},t)=
\frac{1}{\Delta^{2}}\int_{\Delta}d\mbox{\boldmath $\xi$}d\mbox{\boldmath $\xi$}'
\sum_{i=1,j\neq i}^{N/2}\xi'^{b}
\delta(\textbf{r}+\mbox{\boldmath $\xi$}-\textbf{r}_{i}(t))
\delta(\textbf{r}'+\mbox{\boldmath $\xi$}'-\textbf{r}_{j}(t))
\end{equation}
are the two forms of two-particle function of concentration-polarization (\ref{RHD2022ClSCF d 2 def})
or polarization-concentration (\ref{RHD2022ClSCF d 2 def permuted}),
\begin{equation}\label{RHD2022ClSCF n 2 def}
Q_{2}^{bc}(\textbf{r},\textbf{r}',t)=
\frac{1}{\Delta^{2}}\int_{\Delta}d\mbox{\boldmath $\xi$}d\mbox{\boldmath $\xi$}'
\sum_{i=1,j\neq i}^{N/2}\xi^{b}\xi^{c}
\delta(\textbf{r}+\mbox{\boldmath $\xi$}-\textbf{r}_{i}(t))
\delta(\textbf{r}'+\mbox{\boldmath $\xi$}'-\textbf{r}_{j}(t))
\end{equation}
is the two-particle function of concentration-quadrupole moment,
and
\begin{equation}\label{RHD2022ClSCF n 2 def}
D_{2}^{bc}(\textbf{r},\textbf{r}',t)=
\frac{1}{\Delta^{2}}\int_{\Delta}d\mbox{\boldmath $\xi$}d\mbox{\boldmath $\xi$}'
\sum_{i=1,j\neq i}^{N/2}\xi^{b}\xi'^{c}
\delta(\textbf{r}+\mbox{\boldmath $\xi$}-\textbf{r}_{i}(t))
\delta(\textbf{r}'+\mbox{\boldmath $\xi$}'-\textbf{r}_{j}(t))
\end{equation}
is the two-particle polarization-polarization function.

\subsection{Selfconsistent field approximation}


Let us repeat the conclusion about meaning of the self-consistent field approximation.
Our discussion presented above leads to conclusion
that the $\Delta$-vicinity has radius of order of $\sqrt{a\cdot r_{d}}$,
where $a=1/\sqrt[3]{n}$ is the average interparticle distance,
and
$r_{d}$ is the Dabay radius.
The self-consistent field approximation corresponds to regime of interaction of particle being in the nonoverlapping $\Delta$-vicinities.
This condition allows to split the two-particle hydrodynamic functions in the product of corresponding one-particle hydrodynamic functions.

We use the selfconsistent field approximation in equation (\ref{RHD2022ClSCF Phi via n2 d2 Q2}),
so we assume
$n_{2}(\textbf{r},\textbf{r}',t)=n(\textbf{r},t)n(\textbf{r}',t)$,
$d_{2}^{b}(\textbf{r},\textbf{r}',t)=d^{b}(\textbf{r},t)n(\textbf{r}',t)$,
$d_{2}^{b}(\textbf{r}',\textbf{r},t)=n(\textbf{r},t) d^{b}(\textbf{r}',t)$,
$Q_{2}^{bc}(\textbf{r},\textbf{r}',t)=Q^{bc}(\textbf{r},t)n(\textbf{r}',t)$,
$Q_{2}^{bc}(\textbf{r}',\textbf{r},t)=n(\textbf{r},t) Q^{bc}(\textbf{r}',t)$,
and
$D_{2}^{bc}(\textbf{r},\textbf{r}',t)=d^{b}(\textbf{r},t)d^{c}(\textbf{r}',t)$.
Consequently, equation (\ref{RHD2022ClSCF Phi via n2 d2 Q2}) transforms into
$$\Phi^{a}_{e-e}=-q_{e}^{2}\Biggl[
n(\textbf{r},t)\partial^{a}\biggl(\int d\textbf{r}'
G (\mid \textbf{r}-\textbf{r}'\mid)
n(\textbf{r}',t)
-\partial^{b}\int d\textbf{r}'
G (\mid \textbf{r}-\textbf{r}'\mid)
d^{b}(\textbf{r}',t)
+\frac{1}{2}\partial^{b}\partial^{c}\int d\textbf{r}'
G (\mid \textbf{r}-\textbf{r}'\mid)
n(\textbf{r}',t)+...\biggr)$$
\begin{equation}\label{RHD2022ClSCF Phi e-e multipole}
+d^{b}(\textbf{r},t)\partial^{a}\partial^{b}\biggl(\int d\textbf{r}'
G (\mid \textbf{r}-\textbf{r}'\mid)
n(\textbf{r}',t)
-\partial^{c}\int d\textbf{r}'
G (\mid \textbf{r}-\textbf{r}'\mid)
d^{c}(\textbf{r}',t)+...\biggr)
+Q^{bc}(\textbf{r},t)\partial^{a}\partial^{b}\biggl(\int d\textbf{r}'
G (\mid \textbf{r}-\textbf{r}'\mid)
n(\textbf{r}',t)
+...\biggr)
+...\Biggr]
\end{equation}

Equation (\ref{RHD2022ClSCF Phi e-e multipole}) allows to introduce the multipole expansion of the electrostatic potential
\begin{equation}\label{RHD2022ClSCF varphi e}
\varphi_{e}(\textbf{r},t)=q_{e}\int d\textbf{r}'
G (\mid \textbf{r}-\textbf{r}'\mid)
n(\textbf{r}',t)
-\partial^{b}\int d\textbf{r}'
G (\mid \textbf{r}-\textbf{r}'\mid)
d^{b}(\textbf{r}',t)
+\frac{1}{2}\partial^{b}\partial^{c}\int d\textbf{r}'
G (\mid \textbf{r}-\textbf{r}'\mid)
n(\textbf{r}',t)+...\end{equation}
\end{widetext}
Therefore, equation (\ref{RHD2022ClSCF Phi e-e multipole}) reappears structural form
$$\mbox{\boldmath $\Phi$}_{e-e}=q_{e}n_{e}\textbf{E}_{e}$$
\begin{equation}\label{RHD2022ClSCF Phy e-e via E mutipole}
+(\textbf{d}(\textbf{r},t)\cdot\nabla)\textbf{E}_{e}+\frac{1}{2}Q^{bc}(\textbf{r},t)\partial^{b}\partial^{c}\textbf{E}_{e}+...,
\end{equation}
where
$\textbf{E}_{e}=-\nabla\varphi_{e}(\textbf{r},t)$,
and we also can introduce full electric field
$\textbf{E}=\sum_{s=e,i}\textbf{E}_{s}$,
which satisfy the following quasi-static Maxwell equations
$\nabla \times\textbf{E}=0$, and
\begin{equation}\label{RHD2022ClSCF div E}  \nabla \cdot\textbf{E}=4\pi \sum_{s}^{} q_{s}\biggl(n_{s}
+(\nabla\cdot\textbf{d}(\textbf{r},t))
+\frac{1}{2}\partial^{b}\partial^{c}Q^{bc}(\textbf{r},t)
+...\biggr). \end{equation}
Full set of hydrodynamic equations requires the equations for the evolution of for
functions $\textbf{d}(\textbf{r},t)$, $Q^{bc}(\textbf{r},t)$, etc.
We do not present or discuss these equations.
Some information can be found in Ref. \cite{Drofa TMP 96}.

\section{Derivation of the Vlasov equation tracing the microscopic motion of particles}


In order to derive the kinetic theory we need to introduce the distribution function
in the six-dimensional coordinate-momentum space.
We start with the microscopic definition for the system of the point-like particles
\cite{Klimontovich Plasma}
\begin{equation}\label{RHD2022ClSCF distribution function definition zero volume}
f(\textbf{r},\textbf{p},t)=\sum_{i=1}^{N/2}
\delta(\textbf{r}-\textbf{r}_{i}(t))
\delta(\textbf{p}-\textbf{p}_{i}(t)), \end{equation}
where we have the coordinates of particles $\textbf{r}_{i}(t)$,
and their momentums $\textbf{p}_{i}(t)=m_{i}\dot{\textbf{r}}_{i}(t)$
(the nonrelativistic regime).
We use notation $f(\textbf{r},\textbf{p},t)$ for the microscopic distribution function
same as the notation for the macroscopic function below.
However, we can give more detailed representation of arguments of the distribution function
(\ref{RHD2022ClSCF distribution function definition zero volume})
via tracing the time dependence as follows
$f(\textbf{r},\textbf{p},\textbf{r}_{i}(t),\textbf{p}_{i}(t))$
or
$f(\textbf{r},\textbf{p},\{\textbf{r}_{1}(t),\textbf{p}_{1}(t), ..., \textbf{r}_{N/2}(t),\textbf{p}_{N/2}(t)\})$.
Let us repeat that the consideration of the kinetic model of plasmas requires the introduction of the distribution function for each species.
Each distribution function evolves under the action of all species in the system.
We specify
that we consider the quasi-neutral plasmas of two species: the electron-ion plasmas.
We use the following numeration of particles: $i\in [1,N/2]$ for the electrons
and $i\in [N/2+1,N]$ for the ions.
It can be easily represented to the arbitrary set of species in plasmas including neutral particles.
We illustrate the derivation following the evolution of the distribution function of electrons.
The explicit contribution of ions is shown in the terms describing the interaction.

\begin{figure}\includegraphics[width=8cm,angle=0]{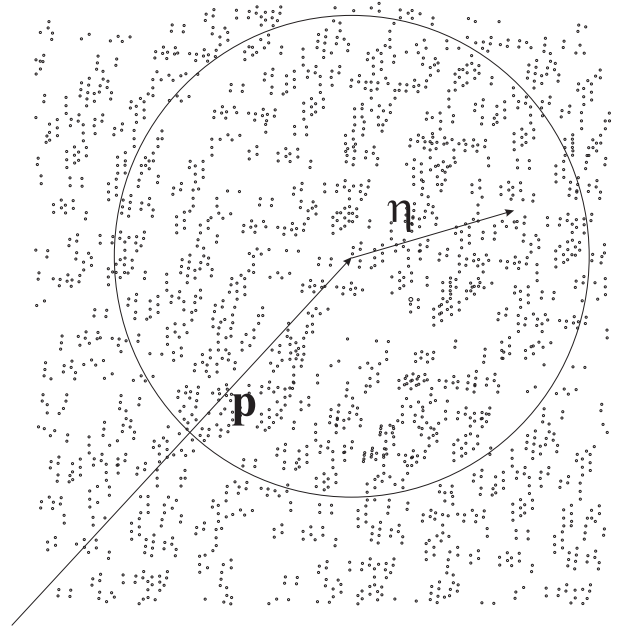}
\caption{\label{RHD2022ClSCF Fig 07}
The consideration of physical kinetics requires analysis of the six-dimensional phase space.
We need to construct the delta-vicinity in the six-dimensional space.
It is basically delta-vicinities in the coordinate space and in the momentum space.
The delta vicinity in the coordinate space is illustrated with Fig. \ref{RHD2022ClSCF Fig 01}.
The delta vicinity in the coordinate space is illustrated here.
Corresponding notations including illustration of vector $\mbox{\boldmath $\eta$}$ scanning the vicinity.}
\end{figure}

Next, we make transition to the physically infinitesimal area.
Moreover, we consider the physically infinitesimal areas both for the coordinate space and the momentum space.
It is constructed in the similar way
as the concentration (\ref{RHD2022ClSCF concentration definition}) presented above
$$f(\textbf{r},\textbf{p},t)=
\frac{1}{\Delta}\frac{1}{\Delta_{p}}\times$$
\begin{equation}\label{RHD2022ClSCF distribution function definition}
\times\int_{\Delta,\Delta_{p}}
d\mbox{\boldmath $\xi$}
d\mbox{\boldmath $\eta$}
\sum_{i=1}^{N/2}
\delta(\textbf{r}+\mbox{\boldmath $\xi$}-\textbf{r}_{i}(t))
\delta(\textbf{p}+\mbox{\boldmath $\eta$}-\textbf{p}_{i}(t)), \end{equation}
where
$d\mbox{\boldmath $\eta$}$ is the element of volume in the momentum space,
$\int_{\Delta,\Delta_{p}}
d\mbox{\boldmath $\xi$}
d\mbox{\boldmath $\eta$}
=\int_{\Delta}d\mbox{\boldmath $\xi$}
\int_{\Delta_{p}}d\mbox{\boldmath $\eta$}$,
with $\int_{\Delta_{p}}d\mbox{\boldmath $\eta$}$ integral over $\Delta_{p}$-vicinity in the momentum space,
$\Delta\equiv \Delta_{r}$ is the delta vicinity in the coordinate space.
As the notion, it is same delta vicinity
which is used above at the derivation of hydrodynamics.
However, its value in kinetics should be larger in order to get the
continuous distribution function.

We need to derive the equations for the evolution of the distribution functions.
Hence, we need to use the equations of motion of the particles.
Here, we consider the nonrelativistic regime for the systems of charged particles.
Consequently, we can use equation (\ref{RHD2022ClSCF Eq of Motion Newton nonRel}).
First, we consider the kinetic equation for the microscopic distribution function
(\ref{RHD2022ClSCF distribution function definition zero volume}).
We take the derivative on time of function (\ref{RHD2022ClSCF distribution function definition zero volume})
and obtain the equation for its evolution
$$\partial_{t}f(\textbf{r},\textbf{p},t)
+\nabla\cdot\sum_{i=1}^{N/2}
\textbf{v}_{i}(t)\delta(\textbf{r}-\textbf{r}_{i}(t))
\delta(\textbf{p}-\textbf{p}_{i}(t))$$
\begin{equation}\label{RHD2022ClSCF kin eq zero volume I}
+\sum_{i=1}^{N/2}
\delta(\textbf{r}-\textbf{r}_{i}(t))
(\dot{\textbf{p}}_{i}(t)\cdot\nabla_{\textbf{p}})\delta(\textbf{p}-\textbf{p}_{i}(t))=0. \end{equation}
We can replace $\nabla_{\textbf{p}}$ to put it in front of whole term,
but it would interfere with the following manipulations with this equation.
We use the equations of motion for each particle.
Using the delta functions
we replace $\textbf{r}_{i}(t)\rightarrow\textbf{r}$
and $\textbf{p}_{i}(t)\rightarrow\textbf{p}$
($\textbf{v}_{i}(t)\rightarrow\textbf{v}=\textbf{p}/m$)
in the expression for the external field acting on $i$-th particle.

This calculation gives the untruncated microscopic kinetic equation
\cite{Klimontovich Plasma}
$$\partial_{t}f(\textbf{r},\textbf{p},t)
+(\textbf{v}\cdot\nabla)f(\textbf{r},\textbf{p},t)$$
$$+\frac{q_{s}}{m_{s}}((\textbf{E}_{ext}(\textbf{r},t)+\textbf{v}\times \textbf{B}_{ext}(\textbf{r},t)/c)
\cdot\nabla_{\textbf{p}})f(\textbf{r},\textbf{p},t)$$
\begin{equation}\label{RHD2022ClSCF kin eq zero volume I}
-\frac{q_{s}q_{s'}}{m_{s}}\nabla_{\textbf{p}}\cdot\sum_{i=1}^{N/2}\sum_{j=1,j\neq i}^{N}
(\nabla_{i}G_{ij})\delta(\textbf{r}-\textbf{r}_{i}(t))
\delta(\textbf{p}-\textbf{p}_{i}(t))=0. \end{equation}
The last term in this equation (\ref{RHD2022ClSCF kin eq zero volume I}) describes the interaction.
The presence of the Green function of the Coulomb interaction leads to the fact that it cannot be expressed via the distribution function,
while it requires the introduction of the two-particle distribution function.

Let us derive the evolution of the distribution function based on exact microscopic motion,
but considered on the scale of the $\Delta$-vicinity (\ref{RHD2022ClSCF distribution function definition}).
We calculate the time derivative of function (\ref{RHD2022ClSCF distribution function definition})
and find the following intermediate equation
\begin{widetext}
$$\partial_{t}f(\textbf{r},\textbf{p},t)+
\nabla\cdot\frac{1}{\Delta}\frac{1}{\Delta_{p}}
\int_{\Delta,\Delta_{p}}
d\mbox{\boldmath $\xi$}
d\mbox{\boldmath $\eta$}
\sum_{i=1}^{N/2} \dot{\textbf{r}}_{i}(t)
\delta(\textbf{r}+\mbox{\boldmath $\xi$}-\textbf{r}_{i}(t))
\delta(\textbf{p}+\mbox{\boldmath $\eta$}-\textbf{p}_{i}(t))$$
\begin{equation}\label{RHD2022ClSCF kin eq delta I}
+\nabla_{\textbf{p}}\cdot\frac{1}{\Delta}\frac{1}{\Delta_{p}}
\int_{\Delta,\Delta_{p}}
d\mbox{\boldmath $\xi$}
d\mbox{\boldmath $\eta$}
\sum_{i=1}^{N/2} \dot{\textbf{p}}_{i}(t)
\delta(\textbf{r}+\mbox{\boldmath $\xi$}-\textbf{r}_{i}(t))
\delta(\textbf{p}+\mbox{\boldmath $\eta$}-\textbf{p}_{i}(t))=0. \end{equation}
The second (third) term appears as the result of action of the time derivative on the delta-function depending on the coordinate (momentum).

Let us make some simple transformations of equation (\ref{RHD2022ClSCF kin eq delta I}).
We use the delta-function depending on the momentum in order to replace the velocity of $i$-th particle $\textbf{v}_{i}(t)=\dot{\textbf{r}}_{i}(t)$
by $(\textbf{p}+\mbox{\boldmath $\eta$})/m_{s}$ ($m_{i}=m_{s}$ for all particles of the species under consideration) in the second term of equation (\ref{RHD2022ClSCF kin eq delta I}).
In the third term of equation (\ref{RHD2022ClSCF kin eq delta I}) we use the equation of motion for $i$-th particle (\ref{RHD2022ClSCF Eq of Motion Newton nonRel}).
Next, we replace the coordinates $\textbf{r}_{i}(t)$ (velocities $\textbf{v}_{i}(t)$)
in the force acting on $i$-th particle by $\textbf{r}+\mbox{\boldmath $\xi$}$ (by $(\textbf{p}+\mbox{\boldmath $\eta$})/m_{s}$).
Hence, we obtain the following representation of equation (\ref{RHD2022ClSCF kin eq delta I}):
$$\partial_{t}f(\textbf{r},\textbf{p},t)+
\frac{1}{m_{s}}\nabla\cdot\frac{1}{\Delta}\frac{1}{\Delta_{p}}
\int_{\Delta,\Delta_{p}}
d\mbox{\boldmath $\xi$}
d\mbox{\boldmath $\eta$}
\sum_{i=1}^{N/2} (\textbf{p}+\mbox{\boldmath $\eta$})
\delta(\textbf{r}+\mbox{\boldmath $\xi$}-\textbf{r}_{i}(t))
\delta(\textbf{p}+\mbox{\boldmath $\eta$}-\textbf{p}_{i}(t))
+\frac{q_{s}}{m_{s}}\frac{1}{\Delta}\frac{1}{\Delta_{p}}
\int_{\Delta,\Delta_{p}}
d\mbox{\boldmath $\xi$}
d\mbox{\boldmath $\eta$}
\sum_{i=1}^{N/2} \times$$
\begin{equation}\label{RHD2022ClSCF kin eq delta II}
\times\biggl(\textbf{E}(\textbf{r}+\mbox{\boldmath $\xi$},t)
+\frac{1}{m_{s}c}[(\textbf{p}+\mbox{\boldmath $\eta$})\times \textbf{B}(\textbf{r}+\mbox{\boldmath $\xi$},t)]
-q_{s'}\nabla_{\textbf{r}}\sum_{j=1,j\neq i}^{N}G(\textbf{r}+\mbox{\boldmath $\xi$}-\textbf{r}_{j}(t))\biggr)
\delta(\textbf{r}+\mbox{\boldmath $\xi$}-\textbf{r}_{i}(t))
\cdot\nabla_{\textbf{p}}\delta(\textbf{p}+\mbox{\boldmath $\eta$}-\textbf{p}_{i}(t))=0. \end{equation}
\end{widetext}
The second term in equation (\ref{RHD2022ClSCF kin eq delta II}) splits on two terms.
The first of them has the well-known form $\textbf{v}\cdot\nabla f_{e}$.
However, the second part of this term has rather
unusual structure
$\frac{1}{m_{s}}\nabla\cdot\frac{1}{\Delta}\frac{1}{\Delta_{p}}
\int_{\Delta,\Delta_{p}}
d\mbox{\boldmath $\xi$}
d\mbox{\boldmath $\eta$}
\sum_{i=1}^{N/2} \mbox{\boldmath $\eta$}
\delta_{\textbf{r}i}
\delta_{\textbf{p}i}\equiv \frac{1}{m_{s}}\nabla\cdot \textbf{f}(\textbf{r},\textbf{p},t)$
related to deviation of the average momentum of particles in the $\Delta_{p}$-vicinity from value $\textbf{p}$ being the center of the vicinity,
where $\delta_{\textbf{r}i}\equiv\delta(\textbf{r}+\mbox{\boldmath $\xi$}-\textbf{r}_{i}(t))$, and
$\delta_{\textbf{p}i}\equiv\delta(\textbf{p}+\mbox{\boldmath $\eta$}-\textbf{p}_{i}(t))$.

The third term in equation (\ref{RHD2022ClSCF kin eq delta II}) contains three terms,
while the second of them presents the Lorentz force.
The Lorentz force $\frac{1}{m_{s}c}[(\textbf{p}+\mbox{\boldmath $\eta$})\times \textbf{B}(\textbf{r}+\mbox{\boldmath $\xi$},t)]$ splits on two terms
due to the deviation of the average momentum $\textbf{p}+\mbox{\boldmath $\eta$}$ from $\textbf{p}$.
Therefore, we have the contribution of the vector distribution function $\textbf{f}(\textbf{r},\textbf{p},t)$ mentioned above.

In addition to the presence of the vector distribution function $\textbf{f}(\textbf{r},\textbf{p},t)$
in kinetic  equation (\ref{RHD2022ClSCF kin eq delta II})
we see necessity to make the multipole expansion of the electric field, the magnetic field, and the Green function of the electron-electron interaction.
Before we make the expansion of the Green function, we need to give symmetric form to term containing this function.
Technical steps are the same as we use for the transformation of the hydrodynamic equations above.
Equation (\ref{RHD2022ClSCF kin eq delta II})
can be rewritten in the following form
\begin{widetext}
$$\partial_{t}f(\textbf{r},\textbf{p},t)+\textbf{v}\cdot\nabla f+\nabla\cdot \textbf{f}(\textbf{r},\textbf{p},t)
+\frac{q_{s}}{m_{s}}\frac{1}{\Delta}\frac{1}{\Delta_{p}}
\int_{\Delta,\Delta_{p}}
d\mbox{\boldmath $\xi$}
d\mbox{\boldmath $\eta$}
\sum_{i=1}^{N/2} \biggl(\textbf{E}(\textbf{r}+\mbox{\boldmath $\xi$},t)
+\frac{1}{m_{s}c}[(\textbf{p}+\mbox{\boldmath $\eta$})\times \textbf{B}(\textbf{r}+\mbox{\boldmath $\xi$},t)]\biggr)
\delta_{\textbf{r}i}
\cdot\nabla_{\textbf{p}}\delta_{\textbf{p}i}$$
\begin{equation}\label{RHD2022ClSCF kin eq delta III}
-\frac{q_{s}}{m_{s}}q_{s'}\nabla_{\textbf{p}}\cdot\frac{1}{\Delta^{2}}\frac{1}{\Delta_{p}^{2}}
\int
d\textbf{r}' d\textbf{p}'
\int_{\Delta,\Delta_{p}}
d\mbox{\boldmath $\xi$}
d\mbox{\boldmath $\eta$}
d\mbox{\boldmath $\xi$}'
d\mbox{\boldmath $\eta$}'
\sum_{i=1}^{N/2}\sum_{j=1, j\neq i}^{N}
\nabla_{\textbf{r}}G(\textbf{r}+\mbox{\boldmath $\xi$}-\textbf{r}'-\mbox{\boldmath $\xi$}')
\delta_{\textbf{r}i}
\delta_{\textbf{p}i}
\delta_{\textbf{r}'j}
\delta_{\textbf{p}'j}
=0, \end{equation}
\end{widetext}
where
$\delta_{\textbf{r}'j}\equiv\delta(\textbf{r}'+\mbox{\boldmath $\xi$}'-\textbf{r}_{j}(t))$, and
$\delta_{\textbf{p}'j}\equiv\delta(\textbf{p}'+\mbox{\boldmath $\eta$}'-\textbf{p}_{j}(t))$.
The fourth and fifth terms describe the action of the external fields and the field of other particle on $i$-th particle
(with further summation on $i$ over all particles).
In this form we cannot include the distribution function in these terms
due to the presence of the electric field, the magnetic field and the Green function under the integral.

The third term in equation (\ref{RHD2022ClSCF kin eq delta III}) contains the vector distribution function,
which has the following explicit form:
$$\textbf{f}(\textbf{r},\textbf{p},t)=
\frac{1}{\Delta}\frac{1}{\Delta_{p}}\times$$
\begin{equation}\label{RHD2022ClSCF distribution VECTOR function definition}
\times\int_{\Delta,\Delta_{p}}
d\mbox{\boldmath $\xi$}
d\mbox{\boldmath $\eta$}
\sum_{i=1}^{N/2}
\mbox{\boldmath $\eta$}
\delta(\textbf{r}+\mbox{\boldmath $\xi$}-\textbf{r}_{i}(t))
\delta(\textbf{p}+\mbox{\boldmath $\eta$}-\textbf{p}_{i}(t)).
\end{equation}



\subsection{Monopole approximation of the kinetic equation}

The dependence of the electric and magnetic fields on $\mbox{\boldmath $\xi$}$
and dependence of the Green function on $\mbox{\boldmath $\xi$}-\mbox{\boldmath $\xi$}'$
do not allow to replace these functions outside of integrals on $d\mbox{\boldmath $\xi$}$
in order to introduce the distribution function in these terms and obtain the closed model.
If functions $E$, $B$, and $G$ slowly change on scale of the delta-vicinity
we can expand these functions on $\mbox{\boldmath $\xi$}$ or $\mbox{\boldmath $\xi$}-\mbox{\boldmath $\xi$}'$, correspondingly.
In this section we consider the first terms in these expansions.
We call it the monopole approximation.


Monopole approximation of equation (\ref{RHD2022ClSCF kin eq delta III}) in coordinate space
(on $\mbox{\boldmath $\xi$}$) of the kinetic equation has the following form:
$$\partial_{t}f(\textbf{r},\textbf{p},t)
+\textbf{v}\cdot\nabla f
+\nabla\cdot \textbf{f}
+\frac{q_{s}}{m_{s}}\biggl(\textbf{E}(\textbf{r},t)
+\frac{1}{c}[\textbf{v}\times \textbf{B}(\textbf{r},t)]\biggr)\cdot\nabla_{\textbf{p}}f$$
$$+\frac{q_{s}}{m_{s}^{2}c}(\nabla_{\textbf{p}}\cdot[\textbf{B}(\textbf{r},t)\times \textbf{f}(\textbf{r},\textbf{p},t)])$$
\begin{equation}\label{RHD2022ClSCF kin eq monopole r}
-\frac{q_{s}}{m_{s}}q_{s'}\nabla_{\textbf{p}}\cdot
\int d\textbf{r}' d\textbf{p}'
(\nabla_{\textbf{r}}G(\textbf{r}-\textbf{r}'))
f_{2}(\textbf{r},\textbf{r}',\textbf{p},\textbf{p}',t)
=0, \end{equation}
where
$q_{s'}f_{2}=q_{e}f_{2,ee}+q_{i}f_{2,ei}$
and
$$f_{2,ee}(\textbf{r},\textbf{r}',\textbf{p},\textbf{p}',t)=$$
\begin{equation}\label{RHD2022ClSCF f 2 def}
=\frac{1}{\Delta^{2}}\frac{1}{\Delta_{p}^{2}}
\int_{\Delta,\Delta_{p}}
d\mbox{\boldmath $\xi$}
d\mbox{\boldmath $\eta$}
d\mbox{\boldmath $\xi$}'
d\mbox{\boldmath $\eta$}'
\sum_{i,j=1, j\neq i}^{N/2}
\delta_{\textbf{r}i}
\delta_{\textbf{p}i}
\delta_{\textbf{r}'j}
\delta_{\textbf{p}'j}
\end{equation}
is the two-particle electron-electron distribution function.

The third and fifth terms should be dropped
if we consider the monopole approximation in the momentum space.

Neglecting $\mbox{\boldmath $\eta$}$ in compare with the momentum $\textbf{p}$ corresponds
to the neglecting of the vector distribution function in equation (\ref{RHD2022ClSCF kin eq monopole r}).


\section{The selfconsistent field approximation in non-relativistic kinetics}

\subsection{Selfconsistent field approximation in the monopole approximation}

Formally, the coordinate part of $\Delta$-vicinity of the point in six-dimensional phase space is introduced in the same way
as it is made for the hydrodynamics.
Above we give an estimation of the radius of the $\Delta_{r}$-vicinity as $\Delta_{r}^{1/3}=\sqrt{a r_{D}}$,
where $a=\sqrt[3]{n}$ is the average interparticle distance,
and $r_{D}$ is the Debay radius.
This physical estimation reflects the mathematical requirement for the construction of the continuous functions
(like the concentration, velocity field, etc)
on the macroscopic scale.
The $\Delta$-vicinity of each point $\textbf{r}$ should contain the large number of particles $N(\textbf{r},t)\gg 1$,
so the change of this number on one or few particles gives the small change of the hydrodynamic functions.
We also have same requirement for the distribution function $f(\textbf{r},\textbf{p},t)$.
But there are stronger restrictions on the number of particles in the six-dimensional $\Delta$-vicinity $N(\textbf{r},\textbf{p},t)$.
Hence, if we have fixed interval of momentum
($\Delta_{p}$-vicinity of point $\textbf{p}$ in the momentum space)
we should have $f(\textbf{r},\textbf{p},t)$ continuous in the coordinate space.
This property should remain at the change of point in the momentum space.
Hence, for each point in the momentum space
we should have
in the coordinate space same number of particles as for the hydrodynamic description $\tilde{N}$.
However, the further summation over all momentum space gives the full number of particles in coordinate $\Delta_{r}$-vicinity up to $\tilde{N}^{2}$.
For the fixed concentration
it increases the scale of $\Delta_{r}$-vicinity in the coordinate space up to $\Delta_{r\in ph}^{1/3}=r_{D}$,
where the subindex $r\in ph$ means that it is coordinate $\Delta_{r}$-vicinity being the part of six-dimensional $\Delta$-vicinity.

To get full the coordinate and momentum monopole approximations
we drop the third and fifth terms in equation (\ref{RHD2022ClSCF kin eq monopole r}).
In order to obtain the present the two-particle distribution function
as the product of the single-particle distribution functions.
Finally, we obtain the Vlasov kinetic equation in the quasi-electrostatic approximation
\begin{equation}\label{RHD2022ClSCF Vlasov eq Coulomb}
\partial_{t}f+\textbf{v}\cdot\nabla f
+q_{s}\biggl(\textbf{E}_{ext}+\textbf{E}+\frac{1}{c}\textbf{v}\times\textbf{B}_{ext}\biggr)
\cdot\frac{\partial f}{\partial \textbf{p}}=0
,\end{equation}
where the electric field is caused by the distribution of charges in the coordinate space:
$\nabla\times \textbf{E}=0$,
and
\begin{equation}\label{RHD2021ClLM div E kin}
\nabla\cdot \textbf{E}=4\pi \sum_{s}^{} q_{s}\int f_{s}(\textbf{r},\textbf{p},t)d\textbf{p}.\end{equation}


The structure of equation for two species electron-ion regime is discussed.
In order to simplify the presentation
we explicitly show the evolution of electrons under the interaction with electrons.
The account of ions can be made in the way described above.

\subsection{Multipole approximation of the kinetic equation}

We use the expansions of the electric field (\ref{RHD2022ClSCF expansion of E}),
the magnetic field (\ref{RHD2022ClSCF expansion of B}),
and the Green function of the Coulomb interaction (\ref{RHD2022ClSCF Green f expansion})
on $\mbox{\boldmath $\xi$}$ and $\mbox{\boldmath $\xi$}-\mbox{\boldmath $\xi$}'$

Equation (\ref{RHD2022ClSCF kin eq monopole r}) appears in the zeroth order multipole expansion in the momentum space.
Here, we consider the multipole expansion of equation (\ref{RHD2022ClSCF kin eq delta III}) in the coordinate space.

We consider terms up to the second order on $\mbox{\boldmath $\xi$}$ or $\mbox{\boldmath $\xi$}-\mbox{\boldmath $\xi$}'$
and find the following equations presented in terms of number of the one-particle distribution functions:
\begin{widetext}
$$\partial_{t}f_{s}+\textbf{v}\cdot\nabla f_{s}+\nabla\cdot \textbf{f}_{s}
+q_{s}\biggl(\textbf{E}_{ext}+\frac{1}{c}\textbf{v}\times\textbf{B}_{ext}\biggr)\cdot\frac{\partial f_{s}}{\partial \textbf{p}}
+q_{s}\biggl(\partial^{b}\textbf{E}_{ext}+\frac{1}{c}\textbf{v}\times\partial^{b}\textbf{B}_{ext}\biggr)
\cdot\frac{\partial d_{s}^{b}(\textbf{r},\textbf{p},t)}{\partial \textbf{p}}$$
$$+q_{e}\biggl(\partial^{b}\partial^{c}\textbf{E}_{ext}
+\frac{1}{c}\textbf{v}\times\partial^{b}\partial^{c}\textbf{B}_{ext}\biggr)\cdot\frac{\partial Q_{s}(\textbf{r},\textbf{p},t)}{\partial \textbf{p}}
+q_{s}\frac{1}{c}\varepsilon^{abc}B_{ext}^{c}\partial_{a,\textbf{p}} f_{s}^{b}
+q_{s}\frac{1}{c}\varepsilon^{abc}\partial^{d}B_{ext}^{c} \partial_{a,\textbf{p}}J_{D,s}^{bd}
+q_{s}\frac{1}{c}\varepsilon^{abc}\partial^{d}\partial^{f}B_{ext}^{c} \partial_{a,\textbf{p}}J_{Q,s}^{bdf}$$
$$-\frac{q_{s}}{m_{s}}\nabla_{\textbf{p}}\cdot
\int d\textbf{r}' d\textbf{p}'
(\nabla_{\textbf{r}}G(\textbf{r}-\textbf{r}'))
f_{2}(\textbf{r},\textbf{r}',\textbf{p},\textbf{p}',t)
-\frac{q_{s}}{m_{s}}\nabla_{\textbf{p}}\cdot
\int d\textbf{r}' d\textbf{p}'
(\nabla_{\textbf{r}}\partial^{b}G(\textbf{r}-\textbf{r}'))
[d_{2}^{b}(\textbf{r},\textbf{r}',\textbf{p},\textbf{p}',t)-d_{2}^{b}(\textbf{r}',\textbf{r},\textbf{p}',\textbf{p},t)] $$
\begin{equation}\label{RHD2022ClSCF Vlasov eq multipole}
-\frac{q_{s}}{m_{s}}\nabla_{\textbf{p}}\cdot
\int d\textbf{r}' d\textbf{p}'
(\nabla_{\textbf{r}}\partial^{b}\partial^{c}G(\textbf{r}-\textbf{r}'))
[Q_{2}^{bc}(\textbf{r},\textbf{r}',\textbf{p},\textbf{p}',t)
+Q_{2}^{bc}(\textbf{r}',\textbf{r},\textbf{p}',\textbf{p},t)
-D_{2}^{bc}(\textbf{r},\textbf{r}',\textbf{p},\textbf{p}',t)
-D_{2}^{cb}(\textbf{r},\textbf{r}',\textbf{p},\textbf{p}',t)]
=0,\end{equation}
\end{widetext}
where
\begin{equation}\label{RHD2022ClSCF d rp def}
d^{a}(\textbf{r},\textbf{p},t)=\frac{1}{\Delta}\frac{1}{\Delta_{p}}
\int_{\Delta,\Delta_{p}}
d\mbox{\boldmath $\xi$}
d\mbox{\boldmath $\eta$}
\sum_{i=1}^{N} \xi^{a}
\delta_{\textbf{r}i}
\delta_{\textbf{p}i}
\end{equation}
is the distribution function of dipole
moment divided by the charge $q_{s}$,
\begin{equation}\label{RHD2022ClSCF Q rp def}
Q^{ab}(\textbf{r},\textbf{p},t)=\frac{1}{\Delta}\frac{1}{\Delta_{p}}
\int_{\Delta,\Delta_{p}} \xi^{a}\xi^{b}
d\mbox{\boldmath $\xi$}
d\mbox{\boldmath $\eta$}
\sum_{i=1}^{N}
\delta_{\textbf{r}i}
\delta_{\textbf{p}i}
\end{equation}
is the distribution function of quadrupole
moment divided by the charge $q_{s}$,
$$d_{2}^{a}(\textbf{r},\textbf{r}',\textbf{p},\textbf{p}',t)=$$
\begin{equation}\label{RHD2022ClSCF d 2 rp def}
=\frac{1}{\Delta^{2}}\frac{1}{\Delta_{p}^{2}}
\int_{\Delta,\Delta_{p}}
d\mbox{\boldmath $\xi$}
d\mbox{\boldmath $\eta$}
d\mbox{\boldmath $\xi$}'
d\mbox{\boldmath $\eta$}'
\sum_{i,j=1, j\neq i}^{N} \xi^{a}
\delta_{\textbf{r}i}
\delta_{\textbf{p}i}
\delta_{\textbf{r}'j}
\delta_{\textbf{p}'j}
\end{equation}
is the two-particle dipole-charge distribution function
divided by the charge $q_{s}$,
$$Q_{2}^{ab}(\textbf{r},\textbf{r}',\textbf{p},\textbf{p}',t)=$$
\begin{equation}\label{RHD2022ClSCF Q 2 rp def}
=\frac{1}{\Delta^{2}}\frac{1}{\Delta_{p}^{2}}
\int_{\Delta,\Delta_{p}}
d\mbox{\boldmath $\xi$}
d\mbox{\boldmath $\eta$}
d\mbox{\boldmath $\xi$}'
d\mbox{\boldmath $\eta$}'
\sum_{i,j=1, j\neq i}^{N} \xi^{a}\xi^{b}
\delta_{\textbf{r}i}
\delta_{\textbf{p}i}
\delta_{\textbf{r}'j}
\delta_{\textbf{p}'j}
\end{equation}
is the two-particle quadrupole-charge distribution function
divided by the charge $q_{s}$,
$$D_{2}^{ab}(\textbf{r},\textbf{r}',\textbf{p},\textbf{p}',t)=$$
\begin{equation}\label{RHD2022ClSCF D 2 rp def}
=\frac{1}{\Delta^{2}}\frac{1}{\Delta_{p}^{2}}
\int_{\Delta,\Delta_{p}}
d\mbox{\boldmath $\xi$}
d\mbox{\boldmath $\eta$}
d\mbox{\boldmath $\xi$}'
d\mbox{\boldmath $\eta$}'
\sum_{i,j=1, j\neq i}^{N} \xi^{a} \xi'^{b}
\delta_{\textbf{r}i}
\delta_{\textbf{p}i}
\delta_{\textbf{r}'j}
\delta_{\textbf{p}'j}
\end{equation}
is the two-particle dipole-dipole distribution function
divided by the charge $q_{s}$,
$$J_{D}^{ab}(\textbf{r},\textbf{p},t)=$$
\begin{equation}\label{RHD2022ClSCF J D rp def}
=\frac{1}{\Delta^{2}}\frac{1}{\Delta_{p}^{2}}
\int_{\Delta,\Delta_{p}}
d\mbox{\boldmath $\xi$}
d\mbox{\boldmath $\eta$}
d\mbox{\boldmath $\xi$}'
d\mbox{\boldmath $\eta$}'
\sum_{i,j=1, j\neq i}^{N} \eta^{a} \xi^{b}
\delta_{\textbf{r}i}
\delta_{\textbf{p}i}
\delta_{\textbf{r}'j}
\delta_{\textbf{p}'j}
\end{equation}
is the distribution function of flux of dipole
moment divided by the charge $q_{s}$,
and
$$J_{Q}^{abc}(\textbf{r},\textbf{p},t)=$$
\begin{equation}\label{RHD2022ClSCF J Q rp def}
=\frac{1}{\Delta^{2}}\frac{1}{\Delta_{p}^{2}}
\int_{\Delta,\Delta_{p}}
d\mbox{\boldmath $\xi$}
d\mbox{\boldmath $\eta$}
d\mbox{\boldmath $\xi$}'
d\mbox{\boldmath $\eta$}'
\sum_{i,j=1, j\neq i}^{N} \eta^{a} \xi^{b} \xi^{c}
\delta_{\textbf{r}i}
\delta_{\textbf{p}i}
\delta_{\textbf{r}'j}
\delta_{\textbf{p}'j}
\end{equation}
is the distribution function of flux of the electric quadrupole moment divided by the charge $q_{s}$.

\subsection{Selfconsistent field approximation in the multipole approximation}

Physically, the selfconsistent field approximation in the multipole regime is the same
as the selfconsistent filed approximation in the monopole regime described above.
Technically, we have splitting of the two-particle distribution functions $d_{2}^{a}$, $Q_{2}^{ab}$, and $D_{2}^{ab}$
on the corresponding one-particle distribution functions.
It gives simplification of kinetic equation (\ref{RHD2022ClSCF Vlasov eq multipole}):
\begin{widetext}
$$\partial_{t}f_{s}+\textbf{v}\cdot\nabla f_{s} +\nabla\cdot \textbf{f}_{s}
+q_{s}\biggl(\textbf{E}_{ext}+\textbf{E}+\frac{1}{c}\textbf{v}\times\textbf{B}_{ext}\biggr)\cdot\frac{\partial f_{s}}{\partial \textbf{p}}
+q_{s}\biggl(\partial^{b}(\textbf{E}_{ext}+\textbf{E})+\frac{1}{c}\textbf{v}\times\partial^{b}\textbf{B}_{ext}\biggr)
\cdot\frac{\partial d_{s}^{b}}{\partial \textbf{p}}$$
\begin{equation}\label{RHD2022ClSCF Vlasov eq multipole SCF appr}
+q_{e}\biggl(\partial^{b}\partial^{c}(\textbf{E}_{ext}+\textbf{E})
+\frac{1}{c}\textbf{v}\times\partial^{b}\partial^{c}\textbf{B}_{ext}\biggr)\cdot\frac{\partial Q_{s}}{\partial \textbf{p}}
+q_{s}\frac{1}{c}\varepsilon^{abc}B_{ext}^{c}\partial_{a,\textbf{p}} f_{s}^{b}
+q_{s}\frac{1}{c}\varepsilon^{abc}\partial^{d}B_{ext}^{c} \partial_{a,\textbf{p}}J_{D,s}^{bd}
+q_{s}\frac{1}{c}\varepsilon^{abc}\partial^{d}\partial^{f}B_{ext}^{c} \partial_{a,\textbf{p}}J_{Q,s}^{bdf}
=0.\end{equation}
\end{widetext}

Kinetic equation (\ref{RHD2022ClSCF Vlasov eq multipole SCF appr})
contains the self-consistent electric field,
which satisfies the following equations:
$\nabla\times \textbf{E}=0$, and
$$\nabla\cdot \textbf{E}=4\pi \sum_{s}^{} q_{s}\biggl(\int f_{s}(\textbf{r},\textbf{p},t)d\textbf{p}$$
\begin{equation}\label{RHD2021ClLM div E kin}
+\partial^{b}\int d_{s}^{b}(\textbf{r},\textbf{p},t)d\textbf{p}
+\frac{1}{2} \partial^{b}\partial^{c}\int Q_{s}^{bc}(\textbf{r},\textbf{p},t)d\textbf{p}
+...\biggr),\end{equation}
where the Coulomb interaction leads to the contribution of the charge dynamics along with the dynamics of the kinetic multipole distribution functions.

Complete analysis of equation  (\ref{RHD2022ClSCF Vlasov eq multipole SCF appr}) requires
the kinetic equations for the additional vector and tensor distribution functions.
These equations can be derived by the method demonstrated in this paper.
But, we do not present these equations here.
This paper is focused on the method of derivation of hydrodynamic and kinetic equations
and on the ways of further generalizations of these models including account of hydrodynamic or kinetic multipole functions.
Examples of closed set of hydrodynamic or kinetic equations consistently including these effects are the subject of further work.


\section{The relativistic
hydrodynamic model with the average reverse gamma factor evolution
and the selfconsistent field approximation in the relativistic hydrodynamics}

Nonrelativistic hydrodynamics and kinetics are considered above.
Basic definitions are given.
The self-consistent field approximation is discussed in terms of suggested model.
Our next goal is the generalization of this method for the systems of relativistic particles.
First, we consider the relativistic hydrodynamics.
We chose the form of relativistic hydrodynamics in the form of
hydrodynamic model with the average reverse gamma factor evolution
recently suggested in Refs. \cite{Andreev 2021 05}, \cite{Andreev 2021 09}.
The hydrodynamic model with the average reverse gamma factor evolution is obtained in the monopole approximation.
Moreover, the interaction between particles and the self-consistent field approximation are not considered explicitly in the cited papers.

The equation of motion for each particle appears as the evolution of the momentum under action of the Lorentz force
$$\dot{\textbf{p}}_{i}(t)=\frac{1}{m_{i}}
\biggl(q_{i}\textbf{E}(\textbf{r}_{i}(t),t)$$
\begin{equation}\label{RHD2022ClSCF Eq of Motion Newton Rel}
+\frac{1}{c}q_{i}[\textbf{v}_{i}(t), \textbf{B}(\textbf{r}_{i}(t),t)]
\biggr), \end{equation}
where
$\textbf{p}_{i}(t)=m_{i}\textbf{v}_{i}(t)/\sqrt{1-\textbf{v}_{i}^{2}(t)/c^{2}}$ is the relativistic momentum,
$\textbf{E}_{i}=\textbf{E}_{i,ext}+\textbf{E}_{i,int}$,
$\textbf{B}_{i}=\textbf{B}_{i,ext}+\textbf{B}_{i,int}$,
$\textbf{E}_{i}=\textbf{E}(\textbf{r}_{i}(t),t)$,
and
$\textbf{B}_{i}=\textbf{B}(\textbf{r}_{i}(t),t)$.
The electric $\textbf{E}_{i,int}$ and magnetic $\textbf{B}_{i,int}$ fields caused by particles surrounding the $i$-th particle
are $\textbf{E}_{i,int}=-\nabla_{i}\varphi(\textbf{r}_{i}(t),t)-\frac{1}{c}\partial_{t}\textbf{A}(\textbf{r}_{i}(t),t)$
and $\textbf{B}_{i,int}=\nabla_{i}\times \textbf{A}(\textbf{r}_{i}(t),t)$
with
\begin{equation}\label{RHD2022ClSCF varphi via Green function rel}
\varphi(\textbf{r}_{i}(t),t)=\sum_{j\neq i}q_{j}\int \frac{\delta(t-t'-\frac{1}{c}\mid \textbf{r}_{i}(t)-\textbf{r}_{j}(t')\mid)}{\mid \textbf{r}_{i}(t)-\textbf{r}_{j}(t')\mid}dt',
\end{equation}
and
\begin{equation}\label{RHD2022ClSCF A via Green function rel}
\textbf{A}(\textbf{r}_{i}(t),t)=\sum_{j\neq i}q_{j}\int \frac{\delta(t-t'-\frac{1}{c}\mid \textbf{r}_{i}(t)-\textbf{r}_{j}(t')\mid)}{\mid \textbf{r}_{i}(t)-\textbf{r}_{j}(t')\mid}
\frac{\textbf{v}_{j}(t')}{c}dt'.
\end{equation}
Here we use
the Green function of the retarding electromagnetic interaction
\begin{equation}\label{RHD2022ClSCF Green function rel}
\tilde{G}_{ij}=
\frac{\delta(t-t'-\frac{1}{c}\mid \textbf{r}_{i}(t)-\textbf{r}_{j}(t')\mid)}{\mid \textbf{r}_{i}(t)-\textbf{r}_{j}(t')\mid}\end{equation}
Here, we demonstrate scalar and vector potentials of electromagnetic field acting on $i$-th particles
and, therefore, created by the surrounding particles.
These potentials (\ref{RHD2022ClSCF varphi via Green function rel})
and (\ref{RHD2022ClSCF A via Green function rel}) can be represented via corresponding Maxwell equations.

Relativistic hydrodynamics is derived by the method described above in Refs. \cite{Andreev 2021 05}, \cite{Andreev 2021 09}.
However, the details of the selfconsistent field approximation is not considered in these papers,
while equations are obtain there in this approximation.

Let us consider the definition of concentration $n$ (\ref{RHD2022ClSCF concentration definition})
for the relativistic regime.
It has same form (\ref{RHD2022ClSCF concentration definition})
which is represented in the shorter form $n=\langle m_{i}\rangle/m$.
We contract the $\Delta$-vicinities in the arbitrary inertial frame.
If we consider the transition to another inertial frame
which moves relatively the first frame with the constant velocity
$\textbf{V}=\{V,0,0\}$
we use the global Lorentz transformation.
All $\Delta$-vicinities change their form due to the contraction of distance in the direction of motion of the second inertial frame:
$\Delta x'=\sqrt{1-V^{2}/c^{2}}\Delta x$.
So, the $\Delta$-vicinities in the second frame are not spherical anymore.
Their volume also change.
However, the number of particles in each vicinity do not change.
So, we have formal change of the concentration at the change of the frame.
Anyway, direct transition of the $\Delta$-vicinities to another frame change their essential property.
They are constructed in the first frame as the motionless $\Delta$-vicinities of the points of space.
While they move in the second frame,
so they are vicinities of moving points.
Hence, the proper formulation of hydrodynamics in the second frame requires
the reconstruction of the $\Delta$-vicinities around motionless point of space.
Therefore, we are focused on the study of the relativistic effects in the fixed frame.
We do not make reference to the "rest frame" since it exists for the relatively simple collective motion of plasmas,
but it does not exist in the general case.

Let us consider the evolution of concentration (\ref{RHD2022ClSCF concentration definition}) in the inertial frame.
We find that it expresses itself via the average velocity of particles
$\textbf{j}=\langle \dot{\textbf{r}}_{i}\rangle=\langle \textbf{v}_{i}\rangle\equiv n\textbf{v}$.
It manifests itself via the continuity equation:
\begin{equation}\label{RHD2022ClSCF cont via v rel hydr} \partial_{t}n+\nabla\cdot(n\textbf{v})=0.\end{equation}

If we want to continue the set of hydrodynamic equations
we need to consider the evolution of current $\textbf{j}$.
Since the current is the average velocity $\langle \textbf{v}_{i}\rangle$
we need to consider the accelerations of all particles.
Therefore, we need to rewrite the equations of motion of each particle
(\ref{RHD2022ClSCF Eq of Motion Newton Rel})
for the velocity change instead of the momentum change:
\begin{equation}\label{RHD2022ClSCF Eq of Motion Newton Rel for acceleration}
\dot{\textbf{v}}_{i}
=\frac{e_{i}}{m_{i}}\sqrt{1-\frac{\textbf{v}_{i}^{2}}{c^{2}}}
\biggl[\textbf{E}_{i}+\frac{1}{c}[\textbf{v}_{i}\times\textbf{B}_{i}]
-\frac{1}{c^{2}}\textbf{v}_{i}(\textbf{v}_{i}\cdot\textbf{E}_{i})\biggr]. \end{equation}

Some further details for derivation of general structure of
hydrodynamic model with the average reverse gamma factor evolution
presented below can be found in Refs. \cite{Andreev 2021 05}, \cite{Andreev 2021 09}.
However, we present the part related to the interaction.

Here, we consider the evolution of the average velocity $\textbf{j}=\langle \textbf{v}_{i}\rangle$:
$$\partial_{t}j^{a}
=\partial_{t}\biggl[\frac{1}{\Delta}\int_{\Delta}d\mbox{\boldmath $\xi$}
\sum_{i=1}^{N} v_{i}^{a}(t) \delta(\textbf{r}+\mbox{\boldmath $\xi$}-\textbf{r}_{i}(t))\biggr]$$
\begin{equation}\label{RHD2022ClSCF j time deriv rel 1}
=-\partial_{b}\Pi^{ab}
+\frac{1}{\Delta}\int_{\Delta}d\mbox{\boldmath $\xi$}
\sum_{i=1}^{N} \dot{v}_{i}^{a}(t) \delta_{i}, \end{equation}
where
\begin{equation}\label{RHD2022ClSCF Pi def rel}
\Pi^{ab}=
\frac{1}{\Delta}\int_{\Delta}d\mbox{\boldmath $\xi$}
\sum_{i=1}^{N} v_{i}^{a}(t)v_{i}^{b}(t) \delta(\textbf{r}+\mbox{\boldmath $\xi$}-\textbf{r}_{i}(t)), \end{equation}
and
$\delta_{i}=\delta(\textbf{r}+\mbox{\boldmath $\xi$}-\textbf{r}_{i}(t)) $.


Usage of equation (\ref{RHD2022ClSCF Eq of Motion Newton Rel for acceleration}) in equation (\ref{RHD2022ClSCF j time deriv rel 1}) gives the following representation of $\partial_{t}j^{a}$:
$$\partial_{t}j^{a} =-\partial_{b}\Pi^{ab}$$
$$+\frac{q_{s}}{m_{s}}\frac{1}{\Delta}\int_{\Delta}d\mbox{\boldmath $\xi$}
\sum_{i=1}^{N}\frac{\delta_{i}}{\gamma_{i}(t)}
\biggl(E^{a}_{i,ext}+\frac{1}{c}\varepsilon^{abc}v_{i}^{b}B_{i,ext}^{c}
-\frac{1}{c^{2}}v_{i}^{a} v_{i}^{b} E^{b}_{i,ext}\biggr)$$
$$+\frac{q_{s}}{m_{s}}\frac{1}{\Delta}\int_{\Delta}d\mbox{\boldmath $\xi$}
\sum_{i=1}^{N}\frac{\delta_{i}}{\gamma_{i}(t)}\Biggl(-\partial^{a}_{i}\varphi_{i,int}-\frac{1}{c}\partial_{t}A_{i,int}^{a}$$
\begin{equation}\label{RHD2022ClSCF j time deriv rel 2}
+\frac{1}{c}\varepsilon_{abc}v_{i}^{b}\varepsilon^{cdf}\partial_{i}^{d}A_{i,int}^{f}
-\frac{1}{c^{2}}v_{i}^{a} v_{i}^{b}\biggl(-\partial^{b}_{i}\varphi_{i,int}-\frac{1}{c}\partial_{t}A_{i,int}^{b}\biggr)\Biggr).
\end{equation}

For the external field we can make the following transformation
using the delta function $\delta_{i}$:
$E^{a}_{i,ext}=E^{a}_{ext}(\textbf{r}_{i}(t),t)=E^{a}_{ext}(\textbf{r}+\mbox{\boldmath $\xi$},t)$
and $B^{a}_{i,ext}=B^{a}_{ext}(\textbf{r}_{i}(t),t)=B^{a}_{ext}(\textbf{r}+\mbox{\boldmath $\xi$},t)$.
So, the external electromagnetic field can be expanded on $\mbox{\boldmath $\xi$}$
if the external electromagnetic field has small changes over the $\Delta$ volume.
Hence, we find
$$F^{a}_{ext}=\frac{q_{s}}{m_{s}}\frac{1}{\Delta}\int_{\Delta}d\mbox{\boldmath $\xi$}
\sum_{i=1}^{N}\frac{\delta_{i}}{\gamma_{i}(t)}\biggl(E^{a}_{i,ext}
+\frac{1}{c}\varepsilon^{abc}v_{i}^{b}B_{i,ext}^{c}$$
$$-\frac{1}{c^{2}}v_{i}^{a} v_{i}^{b} E^{b}_{i,ext}\biggr)
=\frac{q_{s}}{m_{s}}\biggl[ \Gamma E^{a} +\frac{1}{c}\varepsilon^{abc}\Theta^{b}B^{c}-\frac{1}{c^{2}}\Xi^{ab}E^{b}$$
$$+\Gamma_{D}^{b}\partial_{b}E^{a} +\frac{1}{c}\varepsilon^{abc}\Theta_{D}^{bd}\partial_{d}B^{c}-\frac{1}{c^{2}}\Xi_{D}^{abc}\partial_{c}E^{b}$$
\begin{equation}\label{RHD2022ClSCF F a ext expansion rel}
+\Gamma_{Q}^{bc}\partial_{b}\partial_{c}E^{a}+\frac{1}{c}\varepsilon^{abc}\Theta_{Q}^{bdf}\partial_{d}\partial_{f}B^{c}
-\frac{1}{c^{2}}\Xi_{Q}^{abcd}\partial_{c}\partial_{d}E^{b}+...\biggr], \end{equation}
where
the subindex $D$ refers to dipolar,
and the subindex $Q$ refers to quadrupole.
The monopole terms found in Ref. \cite{Andreev 2021 05}
contain the following functions $ \Gamma=\langle \gamma_{i}^{-1}(t)\rangle$,
$\Theta^{b}=\langle \gamma_{i}^{-1}(t)v_{i}^{b}\rangle$,
and
$ \Xi^{ab}=\langle \gamma_{i}^{-1}(t)v_{i}^{a}v_{i}^{b}\rangle$.
The dipolar terms contain the following functions
$ \Gamma_{D}^{b}=\langle \gamma_{i}^{-1}(t)\xi^{b}\rangle$,
$\Theta_{D}^{bd}=\langle \gamma_{i}^{-1}(t)v_{i}^{b}\xi^{d}\rangle$,
and
$ \Xi_{D}^{abc}=\langle \gamma_{i}^{-1}(t)v_{i}^{a}v_{i}^{b}\xi^{c}\rangle$.
While the quadrupolar terms contain the following functions
$ \Gamma_{Q}^{bc}=\langle \gamma_{i}^{-1}(t)\xi^{b}\xi^{b}\rangle$,
$ \Theta_{Q}^{bdf}=\langle \gamma_{i}^{-1}(t)v_{i}^{b}\xi^{d}\xi^{f}\rangle$,
and
$ \Xi_{Q}^{abcd}=\langle \gamma_{i}^{-1}(t)v_{i}^{a}v_{i}^{b}\xi^{c}\xi^{d}\rangle$.

In order to consider the force field of interaction
we start with single term (it is constructed of two first terms)
\begin{equation}\label{RHD2022ClSCF force int 1 01} F^{a}_{int,1}=\frac{q_{s}}{m_{s}}\frac{1}{\Delta}\int_{\Delta}d\mbox{\boldmath $\xi$}
\sum_{i=1}^{N}\frac{\delta_{i}}{\gamma_{i}(t)}E_{i,int}^{a},\end{equation}
where
$E_{i,int}^{a}=-\partial^{a}_{i}\varphi_{i,int}-\frac{1}{c}\partial_{t}A_{i,int}^{a}$.
Using potentials (\ref{RHD2022ClSCF varphi via Green function rel}) and (\ref{RHD2022ClSCF A via Green function rel}) we represent the force via the Green function $G_{ij}$ (\ref{RHD2022ClSCF Green function rel}):
$$F^{a}_{int,1}=-\frac{q_{s}q_{j}}{m_{s}}\frac{1}{\Delta}\int_{\Delta}d\mbox{\boldmath $\xi$}\int dt'
\sum_{i,j=1, i\neq j}^{N}\frac{\delta_{i}}{\gamma_{i}(t)}\times$$
\begin{equation}\label{RHD2022ClSCF force int 1 02}
\times\biggl(\partial_{i}^{a}G_{ij}+\frac{v_{j}^{a}(t')}{c^{2}}\partial_{t}G_{ij}\biggr).\end{equation}
Next, we include additional delta function
$$F^{a}_{int,1}=-\frac{q_{s}q_{j}}{m_{s}}
\frac{1}{\Delta}\int_{\Delta}d\mbox{\boldmath $\xi$}
\int dt'
\int d\textbf{r}'
\sum_{i,j=1, i\neq j}^{N}\frac{1}{\gamma_{i}(t)}\times$$
$$\times\delta(\textbf{r}+\mbox{\boldmath $\xi$}-\textbf{r}_{i}(t))\delta(\textbf{r}'-\textbf{r}_{j}(t'))\times$$
\begin{equation}\label{RHD2022ClSCF force int 1 03}
\times\biggl(\partial_{i}^{a}+\frac{v_{j}^{a}(t')}{c^{2}}\partial_{t}\biggr)G_{ij}(\mid \textbf{r}_{i}(t)-\textbf{r}_{j}(t')\mid,t-t').\end{equation}
We modify the argument of the delta function
$$F^{a}_{int,1}=-\frac{q_{s}q_{j}}{m_{s}}
\frac{1}{\Delta^{2}}\int_{\Delta}d\mbox{\boldmath $\xi$}\int_{\Delta}d\mbox{\boldmath $\xi$}'
\int dt'
\int d\textbf{r}'
\sum_{i,j=1, i\neq j}^{N}$$
$$\frac{1}{\gamma_{i}(t)}
\delta(\textbf{r}+\mbox{\boldmath $\xi$}-\textbf{r}_{i}(t))
\delta(\textbf{r}'+\mbox{\boldmath $\xi$}'-\textbf{r}_{j}(t'))\times$$
\begin{equation}\label{RHD2022ClSCF force int 1 04}
\times\biggl(\partial_{i}^{a}+\frac{v_{j}^{a}(t')}{c^{2}}\partial_{t}\biggr)
G_{ij}(\mid \textbf{r}+\mbox{\boldmath $\xi$}-\textbf{r}-\mbox{\boldmath $\xi$}'\mid,t-t').\end{equation}
At this step we are ready to expand the Green function on $\mbox{\boldmath $\xi$}-\mbox{\boldmath $\xi$}'$
assuming that the Green function has small change over the $\Delta$-vicinity scale:
\begin{widetext}
$$F^{a}_{int,1}=-\frac{q_{s}q_{j}}{m_{s}}
\frac{1}{\Delta^{2}}\int_{\Delta}d\mbox{\boldmath $\xi$}\int_{\Delta}d\mbox{\boldmath $\xi$}'
\int dt'
\int d\textbf{r}'
\sum_{i,j=1, i\neq j}^{N}\frac{1}{\gamma_{i}(t)}
\delta(\textbf{r}+\mbox{\boldmath $\xi$}-\textbf{r}_{i}(t))
\delta(\textbf{r}'+\mbox{\boldmath $\xi$}'-\textbf{r}_{j}(t'))
\biggl(\partial_{i}^{a}+\frac{v_{j}^{a}(t')}{c^{2}}\partial_{t}\biggr)$$
\begin{equation}\label{RHD2022ClSCF force int 1 05}
\biggl(G_{ij}(\mid \textbf{r}-\textbf{r}\mid,t-t')
+(\xi^{b}-\xi'^{b})\partial_{r}^{b}G_{ij}(\mid \textbf{r}-\textbf{r}\mid,t-t')
+\frac{1}{2}(\xi^{b}-\xi'^{b})(\xi^{c}-\xi'^{c})\partial_{r}^{b}\partial_{r}^{c}
G_{ij}(\mid \textbf{r}-\textbf{r}\mid,t-t')+...\biggr). \end{equation}

Let us to interpret the expression presented for the force field
via the two-particle macroscopic functions
$$F^{a}_{int,1}=-\frac{q_{s}q_{j}}{m_{s}}
\int dt'
\int d\textbf{r}'
\Biggl[ \partial_{r}^{a}G \cdot \Gamma_{2}+\frac{1}{c^{2}}\partial_{t}G \cdot X_{1}^{a}$$
$$+\partial_{r}^{a}\partial_{r}^{b}G \cdot X_{2}^{b}+\frac{1}{c^{2}}\partial_{r}^{b}\partial_{t}G \cdot X_{3}^{ab}
-\partial_{r}^{a}\partial_{r}^{b}G \cdot X_{4}^{b}-\frac{1}{c^{2}}\partial_{r}^{b}\partial_{t}G \cdot X_{5}^{ab}
+\partial_{r}^{a}\partial_{r}^{b}\partial_{r}^{c}G \cdot X_{6}^{bc}+\frac{1}{c^{2}}\partial_{r}^{b}\partial_{r}^{c}\partial_{t}G \cdot X_{7}^{abc}$$
\begin{equation}\label{RHD2022ClSCF force int 1 06}
+\partial_{r}^{a}\partial_{r}^{b}\partial_{r}^{c}G \cdot X_{8}^{bc}+\frac{1}{c^{2}}\partial_{r}^{b}\partial_{r}^{c}\partial_{t}G \cdot X_{9}^{abc}
-\partial_{r}^{a}\partial_{r}^{b}\partial_{r}^{c}G \cdot X_{10}^{bc}-\frac{1}{c^{2}}\partial_{r}^{b}\partial_{r}^{c}\partial_{t}G \cdot X_{11}^{abc}
\Biggr],\end{equation}
where $G=G(\textbf{r}-\textbf{r}',t-t')$,
and
we used the following two-particle macroscopic functions,
which are presented together with their limits for the meanfield approximation:
\begin{equation}\label{RHD2022ClSCF def Gamma 2}
\Gamma_{2}(\textbf{r},\textbf{r}',t,t')=
\frac{1}{\Delta^{2}}\int_{\Delta}d\mbox{\boldmath $\xi$}\int_{\Delta}d\mbox{\boldmath $\xi$}'
\sum_{i,j=1, i\neq j}^{N}\frac{1}{\gamma_{i}(t)}
\delta(\textbf{r}+\mbox{\boldmath $\xi$}-\textbf{r}_{i}(t))
\delta(\textbf{r}'+\mbox{\boldmath $\xi$}'-\textbf{r}_{j}(t'))
\equiv \langle\langle\gamma_{i}^{-1}(t)\rangle\rangle \rightarrow \Gamma(\textbf{r},t)\cdot n(\textbf{r}',t'),\end{equation}
\begin{equation}\label{RHD2022ClSCF def X 1} X_{1}^{a}(\textbf{r},\textbf{r}',t,t')
=\langle\langle\gamma_{i}^{-1}(t)\cdot v_{j}^{a}(t')\rangle\rangle \rightarrow \Gamma(\textbf{r},t)\cdot j^{a}(\textbf{r}',t'), \end{equation}
\begin{equation}\label{RHD2022ClSCF def X 2} X_{2}^{b}(\textbf{r},\textbf{r}',t,t')
=\langle\langle\gamma_{i}^{-1}(t)\cdot \rangle\rangle \rightarrow \Gamma_{D}^{b}(\textbf{r},t)\cdot n(\textbf{r}',t'), \end{equation}
\begin{equation}\label{RHD2022ClSCF def X 3} X_{3}^{ab}(\textbf{r},\textbf{r}',t,t')
=\langle\langle\gamma_{i}^{-1}(t)\cdot v_{j}^{a}(t')\rangle\rangle \rightarrow \Gamma_{D}^{b}(\textbf{r},t)\cdot j^{a}(\textbf{r}',t'), \end{equation}
\begin{equation}\label{RHD2022ClSCF def X 4} X_{4}^{b}(\textbf{r},\textbf{r}',t,t')
=\langle\langle\gamma_{i}^{-1}(t)\cdot \rangle\rangle \rightarrow \Gamma(\textbf{r},t)\cdot d^{b}(\textbf{r}',t'), \end{equation}
\begin{equation}\label{RHD2022ClSCF def X 5} X_{5}^{ab}(\textbf{r},\textbf{r}',t,t')
=\langle\langle\gamma_{i}^{-1}(t)\cdot v_{j}^{a}(t')\rangle\rangle \rightarrow \Gamma(\textbf{r},t)\cdot J_{D}^{ab}(\textbf{r}',t'), \end{equation}
\begin{equation}\label{RHD2022ClSCF def X 6} X_{6}^{bc}(\textbf{r},\textbf{r}',t,t')
=\langle\langle\gamma_{i}^{-1}(t)\cdot \rangle\rangle \rightarrow \Gamma_{Q}^{bc}(\textbf{r},t)\cdot n(\textbf{r}',t'), \end{equation}
\begin{equation}\label{RHD2022ClSCF def X 7} X_{7}^{abc}(\textbf{r},\textbf{r}',t,t')
=\langle\langle\gamma_{i}^{-1}(t)\cdot v_{j}^{a}(t')\rangle\rangle \rightarrow \Gamma_{Q}^{bc}(\textbf{r},t)\cdot j^{a}(\textbf{r}',t'), \end{equation}
\begin{equation}\label{RHD2022ClSCF def X 8} X_{8}^{bc}(\textbf{r},\textbf{r}',t,t')
=\langle\langle\gamma_{i}^{-1}(t)\cdot \rangle\rangle \rightarrow \Gamma(\textbf{r},t)\cdot Q^{bc}(\textbf{r}',t'), \end{equation}
\begin{equation}\label{RHD2022ClSCF def X 9} X_{9}^{abc}(\textbf{r},\textbf{r}',t,t')
=\langle\langle\gamma_{i}^{-1}(t)\cdot v_{j}^{a}(t')\rangle\rangle \rightarrow \Gamma(\textbf{r},t)\cdot J_{Q}^{abc}(\textbf{r}',t'), \end{equation}
\begin{equation}\label{RHD2022ClSCF def X 10} X_{10}^{bc}(\textbf{r},\textbf{r}',t,t')
=\langle\langle\gamma_{i}^{-1}(t)\cdot \rangle\rangle \rightarrow \Gamma_{D}^{b}(\textbf{r},t)\cdot d^{c}(\textbf{r}',t'), \end{equation}
and
\begin{equation}\label{RHD2022ClSCF def X 11} X_{11}^{abc}(\textbf{r},\textbf{r}',t,t')
=\langle\langle\gamma_{i}^{-1}(t)\cdot v_{j}^{a}(t')\rangle\rangle \rightarrow \Gamma_{D}^{b}(\textbf{r},t)\cdot J_{D}^{ac}(\textbf{r}',t'). \end{equation}

The force field (\ref{RHD2022ClSCF force int 1 06}) being represented in the selfconsistent field (meanfield) approximation
can be reconstructed as six group of terms
$$F^{a}_{int,1}=
-\frac{q_{s}q_{j}}{m_{s}}\Biggl[
-\Gamma\partial^{a}\Biggl(\int dt'
\int d\textbf{r}' G(\textbf{r},\textbf{r}',t,t') n(\textbf{r}',t')
-\partial^{b}\int dt'
\int d\textbf{r}' G(\textbf{r},\textbf{r}',t,t') d^{b}(\textbf{r}',t')$$
$$+\frac{1}{2}\partial^{b}\partial^{c}\int dt'
\int d\textbf{r}' G(\textbf{r},\textbf{r}',t,t') Q^{bc}(\textbf{r}',t') +...\Biggr)
-\frac{1}{c}\Gamma\cdot\frac{1}{c}\partial_{t}\Biggl(\int dt'
\int d\textbf{r}' G(\textbf{r},\textbf{r}',t,t') j^{a}(\textbf{r}',t')$$
$$
-\partial^{b}\int dt'
\int d\textbf{r}' G(\textbf{r},\textbf{r}',t,t') J_{D}^{ab}(\textbf{r}',t')
+\frac{1}{2}\partial^{b}\partial^{c}\int dt'
\int d\textbf{r}' G(\textbf{r},\textbf{r}',t,t') J_{Q}^{abc}(\textbf{r}',t') +...\Biggr)$$
$$-\Gamma_{D}^{b}\partial^{a}\partial^{b}\Biggl(\int dt'
\int d\textbf{r}' G(\textbf{r},\textbf{r}',t,t') n(\textbf{r}',t')
-\partial^{b}\int dt'
\int d\textbf{r}' G(\textbf{r},\textbf{r}',t,t') d^{c}(\textbf{r}',t') +...\Biggr)$$
$$-\frac{1}{c}\Gamma_{D}^{b}\cdot\frac{1}{c}\partial^{b}\partial_{t}\Biggl(\int dt'
\int d\textbf{r}' G(\textbf{r},\textbf{r}',t,t') j^{a}(\textbf{r}',t')
-\partial^{c}\int dt'
\int d\textbf{r}' G(\textbf{r},\textbf{r}',t,t') J_{D}^{ac}(\textbf{r}',t') +...\Biggr)$$
\begin{equation}\label{RHD2022ClSCF force int 1 07}
-\frac{1}{2}\Gamma_{Q}^{bc}\partial^{a}\partial^{b}\partial^{c}\Biggl(\int dt'
\int d\textbf{r}' G(\textbf{r},\textbf{r}',t,t') n(\textbf{r}',t') +...\Biggr)
-\frac{1}{2}\frac{1}{c}\Gamma_{Q}^{bc}\cdot\frac{1}{c}\partial_{t}\partial^{b}\partial^{c}\Biggl(\int dt'
\int d\textbf{r}' G(\textbf{r},\textbf{r}',t,t') j^{a}(\textbf{r}',t') +...\Biggr)
\Biggr]. \end{equation}
These structures allows to introduce the macroscopic scalar and vector potentials
\begin{equation}\label{RHD2022ClSCF varphi macro exp}
\varphi(\textbf{r},t)=\int dt'
\int d\textbf{r}' G(\textbf{r},\textbf{r}',t,t') n(\textbf{r}',t')
-\partial^{b}\int dt'
\int d\textbf{r}' G(\textbf{r},\textbf{r}',t,t') d^{b}(\textbf{r}',t')
+\frac{1}{2}\partial^{b}\partial^{c}\int dt'
\int d\textbf{r}' G(\textbf{r},\textbf{r}',t,t') Q^{bc}(\textbf{r}',t') +... ,\end{equation}
and
\begin{equation}\label{RHD2022ClSCF A macro exp}
A^{a}(\textbf{r},t)=\int dt' \biggl(
\int d\textbf{r}' G(\textbf{r},\textbf{r}',t,t') j^{a}(\textbf{r}',t')
-\partial^{b}
\int d\textbf{r}' G(\textbf{r},\textbf{r}',t,t') J_{D}^{ab}(\textbf{r}',t')
+\frac{1}{2}\partial^{b}\partial^{c}
\int d\textbf{r}' G(\textbf{r},\textbf{r}',t,t') J_{Q}^{abc}(\textbf{r}',t') +... \biggr).\end{equation}
\end{widetext}
These potentials allow to introduce the macroscopic electric and magnetic fields
$\textbf{E}=-\nabla\varphi(\textbf{r},t)-(1/c)\partial_{t}\textbf{A}$
and
$\textbf{B}=curl \textbf{A}$.
Considered part of the force field is expressed via the electric field only
\begin{equation}\label{RHD2022ClSCF force int 1 08} F^{a}_{int,1}=
-\frac{q_{s}q_{j}}{m_{s}} (\Gamma E^{a}
+\Gamma_{D}^{b}\partial^{b}E^{a}+\frac{1}{2}\Gamma_{Q}^{bc}\partial^{b}\partial^{c}E^{a}).\end{equation}
While electromagnetic this field,
in accordance with the explicit form of the Green function (\ref{RHD2022ClSCF Green function rel}),
satisfies the Maxwell equations
\begin{equation}\label{RHD2022ClSCF div B multi} \nabla \cdot\textbf{B}=0, \end{equation}
\begin{equation}\label{RHD2022ClSCF rot E multi with time}
\nabla\times \textbf{E}=-\frac{1}{c}\partial_{t}\textbf{B},  \end{equation}
\begin{equation}\label{RHD2022ClSCF div E multi}
\nabla \cdot\textbf{E}=\sum_{s}4\pi q_{s}\biggl(n_{s}-\nabla\cdot \textbf{d}_{s}+\frac{1}{2}\partial_{a}\partial_{b}Q_{s}^{ab}+...\biggr), \end{equation}
and
$$(\nabla\times \textbf{B})^{a}=\frac{1}{c}\partial_{t}E^{a}$$
\begin{equation}\label{RHD2022ClSCF rot B multi with time}
+\sum_{s}\frac{4\pi q_{s}}{c}\biggl(n_{s}v_{s}^{a}-\partial_{b}J^{ab}_{D,s} +\frac{1}{2}\partial_{b}\partial_{c}J_{Q,s}^{abc} +...\biggr) ,\end{equation}
where the Lorentz Gauge is used $\partial_{t}\varphi/c+\nabla \textbf{A}=0$.

Other group of terms in equation (\ref{RHD2022ClSCF j time deriv rel 2}) can be presented
via the macroscopic scalar and vector potentials in similar way.

In the monopole approximation of the hydrodynamic model with the average reverse gamma factor evolution
we have the following equation of evolution of the velocity field
$$n\partial_{t}v^{a}+n(\textbf{v}\cdot\nabla)v^{a}+\partial^{a}\tilde{p}
=\frac{q}{m}\Gamma E^{a}+\frac{q}{mc}\varepsilon^{abc}(\Gamma v^{b}+t^{b})B^{c}$$
\begin{equation}\label{RHD2022ClSCF Euler for v rel hydr} -\frac{q}{mc^{2}}(\Gamma v^{a} v^{b}+v^{a}t^{b}+v^{b}t^{a})E^{b}
-\frac{e}{mc^{2}}\tilde{t}E^{a}, \end{equation}
where
$\Gamma=\langle \frac{1}{\gamma_{i}}\rangle$,
$t^{a}=\langle \frac{1}{\gamma_{i}}v_{i}^{a} \rangle -\Gamma v^{a}$,
$p^{ab}=\langle v_{i}^{a}v_{i}^{b} \rangle-n v^{a}v^{b}$,
$t^{ab}=\langle \frac{1}{\gamma_{i}}v_{i}^{a}v_{i}^{b} \rangle-\Gamma v^{a}v^{b}-t^{a}v^{b}- v^{a}t^{b}$,
$\gamma_{i}=1/\sqrt{1-\textbf{v}_{i}(t)^{2}/c^{2}}$.
Two equations of state should be applied for functions $\tilde{p}$ ($p^{ab}=\tilde{p}\delta^{ab}$)
and $\tilde{t}$ ($t^{ab}=\tilde{t}\delta^{ab}$) (see Ref. \cite{Andreev 2021 05}).
We also have corresponding simplification of the Maxwell equations
(\ref{RHD2022ClSCF div B multi}), (\ref{RHD2022ClSCF rot E multi with time}),
(\ref{RHD2022ClSCF div E multi}), (\ref{RHD2022ClSCF rot B multi with time}):
$ \nabla \cdot\textbf{B}=0$,
\begin{equation}\label{RHD2022ClSCF rot E and div E} \begin{array}{cc}
\nabla\times \textbf{E}=-\frac{1}{c}\partial_{t}\textbf{B}, & \nabla \cdot\textbf{E}=4\pi(en_{i}-en_{e}),\end{array} \end{equation}
\begin{equation}\label{RHD2022ClSCF rot B with time}
\nabla\times \textbf{B}=\frac{1}{c}\partial_{t}\textbf{E}+\sum_{s}\frac{4\pi q_{s}}{c}n_{s}\textbf{v}_{s}.\end{equation}


Next, the hydrodynamic model with the average reverse gamma factor evolution suggests
the derivation of the equation for evolution of the average reverse gamma factor $\Gamma$.
This derivation is similar to the derivation presented above for the particle current evolution.
It also includes the contribution of the multipole moments.
We do not show this derivation assuming
that the illustration made for the particle current evolution is enough for the purpose of this paper.
The equation for evolution of the average reverse gamma factor $\Gamma$
has the following form in accordance with Refs. \cite{Andreev 2021 05}, \cite{Andreev 2021 09}
\begin{equation}\label{RHD2022ClSCF eq for Gamma} \partial_{t}\Gamma+\partial_{b}(\Gamma v^{b}+t^{b})
=-\frac{q}{mc^{2}}n\textbf{v}\cdot\textbf{E}\biggl(1-\frac{1}{c^{2}}\biggl(\textbf{v}^{2}+\frac{5p}{n}\biggr)\biggr).\end{equation}

The forth and final equation for the evolution of the material field in
the hydrodynamic model with the average reverse gamma factor evolution
is the evolution of the flux of the average reverse gamma factor \cite{Andreev 2021 05}, \cite{Andreev 2021 09}:
$$(\partial_{t}+\textbf{v}\cdot\nabla)t^{a}+\partial_{a}\tilde{t}
+(\textbf{t}\cdot\nabla) v^{a}+t^{a} (\nabla\cdot \textbf{v})$$
$$+\Gamma(\partial_{t}+\textbf{v}\cdot\nabla)v^{a}
=\frac{q}{m}nE^{a}\biggl[1-\frac{\textbf{v}^{2}}{c^{2}}-\frac{3p}{nc^{2}}\biggr]$$
$$+\frac{q}{mc}\varepsilon^{abc}nv^{b}B^{c}\biggl[1-\frac{\textbf{v}^{2}}{c^{2}}-\frac{5p}{nc^{2}}\biggr]
-\frac{2q}{mc^{2}}E^{a}p\biggl[1-\frac{\textbf{v}^{2}}{c^{2}}\biggr]$$
\begin{equation}\label{RHD2022ClSCF eq for t a} -\frac{q}{mc^{2}}nv^{a}v^{b}E^{b}\biggl[1-\frac{\textbf{v}^{2}}{c^{2}}-\frac{9p}{nc^{2}}\biggr]
+\frac{10q}{3mc^{4}}\tilde{M} E^{a}.\end{equation}
Function $\tilde{M}$ appears as the simplification for the fourth rank tensor
$M^{abcd}=\frac{\tilde{M}}{3}(\delta^{ab}\delta^{cd}+\delta^{ac}\delta^{bd}+\delta^{ad}\delta^{bc})$
see equation (17) of Ref. \cite{Andreev 2021 05}.
Equation (\ref{RHD2022ClSCF eq for t a}) shows that we need to include the third equation of state
for the function $\tilde{M}$.
It is found in Ref. \cite{Andreev 2021 05}.

\section{The selfconsistent field approximation in the relativistic Vlasov equation}

Above we present  the derivation of the kinetic equation in the nonrelativistic regime,
where the interaction is restricted by the Coulomb interaction.
However, the contribution of the multipole moments is considered as well.
Here, we consider the kinetic model for the relativistic motion particles
(\ref{RHD2022ClSCF Eq of Motion Newton Rel}).
Moreover, we assume
that the interaction between particles is the full electromagnetic interaction,
so the field acting on $i$-th particle is created by surrounding particles in accordance with the microscopic Maxwell equations
(see equations (\ref{RHD2022ClSCF Eq of Motion Newton Rel}),
(\ref{RHD2022ClSCF varphi via Green function rel}), (\ref{RHD2022ClSCF A via Green function rel})).
We use the definition (\ref{RHD2022ClSCF distribution function definition}) for the distribution function.
Hence, its time derivative has form of equation (\ref{RHD2022ClSCF kin eq delta I}).
Usage of equation (\ref{RHD2022ClSCF Eq of Motion Newton Rel})
for the time derivative of the momentum of $i$-th particle gives the following equation
\begin{widetext}
$$\partial_{t}f(\textbf{r},\textbf{p},t)
+\nabla\cdot \textbf{F}(\textbf{r},\textbf{p},t)
+\frac{q_{s}}{m_{s}}\nabla_{\textbf{p}}\cdot\frac{1}{\Delta}\frac{1}{\Delta_{p}}
\int_{\Delta,\Delta_{p}}
d\mbox{\boldmath $\xi$}
d\mbox{\boldmath $\eta$}
\sum_{i=1}^{N/2} \biggl(\textbf{E}_{ext}(\textbf{r}+\mbox{\boldmath $\xi$},t)
+\frac{1}{c}[\textbf{v}_{i}(t)\times \textbf{B}_{ext}(\textbf{r}+\mbox{\boldmath $\xi$},t)]\biggr)
\delta_{\textbf{r}i}
\delta_{\textbf{p}i}$$
$$-\frac{q_{s}}{m_{s}}q_{s'}\nabla_{\textbf{p}}\cdot\frac{1}{\Delta^{2}}\frac{1}{\Delta_{p}^{2}}
\int dt'
\int
d\textbf{r}' d\textbf{p}'
\int_{\Delta,\Delta_{p}}
d\mbox{\boldmath $\xi$}
d\mbox{\boldmath $\eta$}
d\mbox{\boldmath $\xi$}'
d\mbox{\boldmath $\eta$}'
\sum_{i=1}^{N/2}\sum_{j=1, j\neq i}^{N}
\delta_{\textbf{r}i}
\delta_{\textbf{p}i}
\delta_{\textbf{r}'j}
\delta_{\textbf{p}'j}\times$$
\begin{equation}\label{RHD2022ClSCF kin eq delta III-REL}
\times\biggl(
\biggl(1-\frac{\textbf{v}_{i}(t)\cdot\textbf{v}_{j}(t')}{c^{2}}\biggr)
\nabla_{\textbf{r}}G(\textbf{r}+\mbox{\boldmath $\xi$}-\textbf{r}'-\mbox{\boldmath $\xi$}')
+\frac{\textbf{v}_{j}(t')}{c^{2}}(\partial_{t}+(\textbf{v}_{i}(t)\cdot\nabla_{\textbf{r}})G(\textbf{r}+\mbox{\boldmath $\xi$}-\textbf{r}'-\mbox{\boldmath $\xi$}')\biggr)
=0,
\end{equation}
where
$\delta_{\textbf{r}'j}\equiv\delta(\textbf{r}'+\mbox{\boldmath $\xi$}'-\textbf{r}_{j}(t))$, and
$\delta_{\textbf{p}'j}\equiv\delta(\textbf{p}'+\mbox{\boldmath $\eta$}'-\textbf{p}_{j}(t))$.
Equation (\ref{RHD2022ClSCF kin eq delta III-REL}) contains function
\begin{equation}\label{RHD2022ClSCF function F def REL}
\textbf{F}(\textbf{r},\textbf{p},t)=\frac{1}{\Delta}\frac{1}{\Delta_{p}}
\int_{\Delta,\Delta_{p}}
d\mbox{\boldmath $\xi$}
d\mbox{\boldmath $\eta$}
\sum_{i=1}^{N/2} \textbf{v}_{i}(t)
\delta_{\textbf{r}i}
\delta_{\textbf{p}i}.
\end{equation}
We can use relativistic expression for the velocity via momentum
$\textbf{v}_{i}(t)=\textbf{p}_{i}(t)c/\sqrt{\textbf{p}_{i}^{2}(t)+m_{i}^{2}c^{2}}$
with further replacement of the momentum $\textbf{p}_{i}(t)$ on $\textbf{p}+\mbox{\boldmath $\eta$}$.
Hence function $\textbf{F}(\textbf{r},\textbf{p},t)$ contains nonpolinomic dependence on $\mbox{\boldmath $\eta$}$:
\begin{equation}\label{RHD2022ClSCF function F repr 1 REL}
\textbf{F}(\textbf{r},\textbf{p},t)=\frac{1}{\Delta}\frac{1}{\Delta_{p}}
\int_{\Delta,\Delta_{p}}
d\mbox{\boldmath $\xi$}
d\mbox{\boldmath $\eta$}
\sum_{i=1}^{N/2} \frac{(\textbf{p}+\mbox{\boldmath $\eta$})c}{\sqrt{(\textbf{p}+\mbox{\boldmath $\eta$})^{2}+m_{i}^{2}c^{2}}}
\delta_{\textbf{r}i}
\delta_{\textbf{p}i}.
\end{equation}
Same replacement should be made in for the velocities in other terms in equation (\ref{RHD2022ClSCF kin eq delta III-REL}).

Let us consider expansion on $\mbox{\boldmath $\eta$}$ as the small value in compare with $\textbf{p}$ and $m_{i}c$:
\begin{equation}\label{RHD2022ClSCF function F repr 2 REL}
\textbf{F}(\textbf{r},\textbf{p},t)=\textbf{v}\cdot f(\textbf{r},\textbf{p},t)+\frac{1}{\Delta}\frac{1}{\Delta_{p}}
\int_{\Delta,\Delta_{p}}
d\mbox{\boldmath $\xi$}
d\mbox{\boldmath $\eta$}
\sum_{i=1}^{N/2}
\biggl(\frac{\mbox{\boldmath $\eta$}c}{\sqrt{\textbf{p}^{2}+m_{i}^{2}c^{2}}}
-\frac{\textbf{p}(\textbf{p}\cdot\mbox{\boldmath $\eta$})c}{(\sqrt{\textbf{p}^{2}+m_{i}^{2}c^{2}})^{3}}\biggr)
\delta_{\textbf{r}i}
\delta_{\textbf{p}i}.
\end{equation}
If we consider the monopole regime in the momentum space
we neglect $\mbox{\boldmath $\eta$}$
and find
$\textbf{F}(\textbf{r},\textbf{p},t)=\textbf{v}\cdot f(\textbf{r},\textbf{p},t)$.
Here, we also use $\textbf{p}=m_{s}\textbf{v}/\sqrt{1-\textbf{v}^{2}/c^{2}}$ and
$\textbf{v}=\textbf{p}c/\sqrt{\textbf{p}^{2}+m_{s}^{2}c^{2}}$.

We consider this equation in the monopole approximation.
Hence, equation (\ref{RHD2022ClSCF kin eq delta III-REL}) simplifies to
$$\partial_{t}f(\textbf{r},\textbf{p},t)+\textbf{v}\cdot\nabla f
+\frac{q_{s}}{m_{s}}\frac{1}{\Delta}\frac{1}{\Delta_{p}}
\int_{\Delta,\Delta_{p}}
d\mbox{\boldmath $\xi$}
d\mbox{\boldmath $\eta$}
\sum_{i=1}^{N/2} \biggl(\textbf{E}_{ext}(\textbf{r},t)
+\frac{1}{c}[(\textbf{v})\times \textbf{B}_{ext}(\textbf{r},t)]\biggr)
\delta_{\textbf{r}i}
\cdot\nabla_{\textbf{p}}\delta_{\textbf{p}i}$$

$$-\frac{q_{s}}{m_{s}}q_{s'}\cdot\frac{1}{\Delta^{2}}\frac{1}{\Delta_{p}^{2}}
\int dt'
\int
d\textbf{r}' d\textbf{p}'
\times$$
\begin{equation}\label{RHD2022ClSCF kin eq delta IV-REL}
\times\biggl(
\biggl(1-\frac{\textbf{v}\cdot\textbf{v}'}{c^{2}}\biggr)
\nabla_{\textbf{r}}G(\textbf{r}-\textbf{r}', t-t')
+\frac{\textbf{v}'}{c^{2}}(\partial_{t}+(\textbf{v}\cdot\nabla_{\textbf{r}}))G(\textbf{r}-\textbf{r}', t-t')\biggr)
\nabla_{\textbf{p}}f_{2}(\textbf{r},\textbf{p},t, \textbf{r}',\textbf{p}',t')
=0,
\end{equation}
where two-particle distribution function is presented in accordance with definition (\ref{RHD2022ClSCF f 2 def}),
but including dependence on $t'$:
\begin{equation}\label{RHD2022ClSCF f 2 def REL} f_{2}(\textbf{r},\textbf{p},t, \textbf{r}',\textbf{p}',t')=\int_{\Delta,\Delta_{p}}
d\mbox{\boldmath $\xi$}
d\mbox{\boldmath $\eta$}
d\mbox{\boldmath $\xi$}'
d\mbox{\boldmath $\eta$}'
\sum_{i=1}^{N/2}\sum_{j=1, j\neq i}^{N}
\delta_{\textbf{r}i}
\delta_{\textbf{p}i}
\delta_{\textbf{r}'j}
\delta_{\textbf{p}'j}.
\end{equation}

Here, we make transition to the electromagnetic field instead of integral form of the kinetic equation
\begin{equation}\label{RHD2022ClSCF kin eq delta IV-REL}\partial_{t}f(\textbf{r},\textbf{p},t)+\textbf{v}\cdot\nabla f
+\frac{q_{s}}{m_{s}}\frac{1}{\Delta}\frac{1}{\Delta_{p}}
\int_{\Delta,\Delta_{p}}
d\mbox{\boldmath $\xi$}
d\mbox{\boldmath $\eta$}
\sum_{i=1}^{N/2} \biggl(\textbf{E}(\textbf{r},t)
+\frac{1}{c}[\textbf{v}\times \textbf{B}(\textbf{r},t)]\biggr)
\delta_{\textbf{r}i}
\cdot\nabla_{\textbf{p}}\delta_{\textbf{p}i}=0,\end{equation}
\end{widetext}
where $\textbf{E}(\textbf{r},t)$ and $\textbf{B}(\textbf{r},t)$
are the full fields composed of the external fields (like $\textbf{E}_{ext}(\textbf{r},t)$)
and the fields of interparticle interaction (like $\textbf{E}_{int}(\textbf{r},t)$)
$\textbf{E}(\textbf{r},t)=\textbf{E}_{ext}(\textbf{r},t)+\textbf{E}_{int}(\textbf{r},t)$
and $\textbf{B}(\textbf{r},t)=\textbf{B}_{ext}(\textbf{r},t)+\textbf{B}_{int}(\textbf{r},t)$.
While the field of interaction satisfies the Maxwell equations:
$ \nabla \cdot\textbf{B}_{int}=0$,
\begin{equation}\label{RHD2022ClSCF rot E and div E} \begin{array}{cc}
\nabla\times \textbf{E}_{int}=-\frac{1}{c}\partial_{t}\textbf{B}_{int}, &
\nabla \cdot\textbf{E}_{int}=4\pi\sum_{s}q_{s}\int f(\textbf{r},\textbf{p},t)d\textbf{p},\end{array} \end{equation}
\begin{equation}\label{RHD2022ClSCF rot B with time}
\nabla\times \textbf{B}_{int}=\frac{1}{c}\partial_{t}\textbf{E}_{int}+\frac{4\pi}{c}\sum_{s}q_{s}\int \textbf{v} f(\textbf{r},\textbf{p},t)d\textbf{p}.\end{equation}


Finally, the full set of the Vlasov-Maxwell equations is obtained in the selfconsistent field approximation for the relativistic plasmas.




\section{Conclusion}

There has been an open problem of derivation of the Vlason kinetic equation for the full relativistic regime,
where the electromagnetic field created by each particle satisfies the full set of Maxwell equations
along with the possibility of high temperatures.
Similar problem has been considered for the relativistic hydrodynamics.
Both problems have been addressed in this paper and the method of solving of these problems has been demonstrated explicitly,
including all essential technical details.

In order to simplify the presentation, the paper has been splitted on several sections.
Firstly, the nonrelativistic hydrodynamics has been considered.
So, the interparticle interaction has been reduced to the Coulomb interaction.
It has been derived from the microscopic motion of classic particles.
All paper has been focused on the classic systems with no discussion of quantum effects.
The presented method of derivation includes the transition on the macroscopic scale from the microscopic level of description.
Hence, the physically infinitesimal volume has been presented analytically.
Dealing with the microscopic description
we consider the finite elements of volume as the macroscopically point-like objects.
However, these volumes (which are called $\Delta$-vicinity) are characterized by density of the dipole moment,
the density of quadrupole moment, etc, in addition to the charge density.
The contribution of the multipole moments in the Euler equation and the Poisson equation has been illustrated.
The equations for the multipole moments evolution are not presented since our goal in this paper is to give background for the further derivation of the relativistic hydrodynamic and kinetic models.
Hence, the features related to the multipole moments have been mentioned,
but no stress has been maid on this item.
The selfconsistent field approximation (the mean field approximation) has been discussed for the nonrelativistic plasmas as well.

The second part of this paper has been focused to the derivation of the Vlasov kinetic equation, in the nonrelativistic regime,
where particles interact via the Coulomb interaction.
The derivation includes the transition on the macroscopic scale both in the coordinate and momentum space.
Hence, the distribution function for the multipole moments have been found.
No additional kinetic equations have been considered for these functions,
but their presence in the Vlasov equation and the Poisson equation has been highlighted.
Neglecting the multipole expansion of the Coulomb interaction
we still have additional vector distribution function.
It can be characterized as the "dipole moment of the vicinity in the momentum space".
If we neglect the contribution of all additional functions
we find the kinetic equation with the two-particle distribution function.
Further application of the self-consistent field approximation leads to the well-known Vlasov equation in the Coulomb approximation.

The third item is one of two major items in this paper.
It is the derivation of the relativistic hydrodynamics.
Let us repeat that the suggested method of derivation of hydrodynamic and kinetic models
shows the method of explicit transition to the macroscopic scale.
It has been called the method of averaging,
but it has no statistical or probabilistic meaning.
While it gives transition of deterministic behavior on another scale of space parameters.
Suggested method allows to derive the relativistic hydrodynamics in
the well-known form of set of continuity equation and four momentum evolution equation
(see for instance Ref. \cite{Andreev 2021 09}).
So, this model will include the multipole expansion demonstrated in this paper.
However, another form relativistic hydrodynamic model is chosen for the analysis.
It is the relativistic hydrodynamic model with the average reverse gamma factor evolution.
The continuity equation and the equation for the evolution of the current of particles,
which transforms to the velocity field evolution equation are derived in full details.
While the evolution of the average reverse gamma factor and its current demonstrate similar structures,
so them have not been demonstrated.
The derivation includes two essential elements.
First, it is the relativistic temperatures of the plasmas.
Second, it is full relativistic interaction between particles,
so the electromagnetic field created by each particle satisfies the full set of Maxwell equations.


The fourth item is the full relativistic derivation of the Vlasov kinetic equation.
This elements of the paper is major part from the fundamental point of view.
The derivation itself has been presented via few equations.
From technical point of view
it is similar to the derivation of the nonrelativistic version.
But, it is essential
that described method allows to give derivation at the relativistically large temperatures
while the interparticle interaction happens in accordance with
the full set of Maxwell equations at the microscopic level.
So, the selfconsistent field macroscopic field also satisfies the full set of Maxwell equations.
Moreover, this derivation is rather simple  and straightforward.
So, it is easy for understanding.

Described derivation includes some open problems for the hydrodynamics and kinetics.
Major problem is the complete analysis of the multipole moments and their dynamics.
Corresponding equations for theevolution of these functions are not discussed.
Moreover, the closed set of equations consistently
describing these effects especially in kinetics is to be found.



\section{DATA AVAILABILITY}

Data sharing is not applicable to this article as no new data were
created or analyzed in this study, which is a purely theoretical one.




\begin{thebibliography}{17}


\bibitem{Weinberg Gr 72} S. Weinberg, Gravitation and Cosmology (John Wiley and Sons, Inc., NewYork, 1972).





\bibitem{Vlasov JETP 38} A. A. Vlasov,  J. Exp. Theor. Phys. \textbf{8},  291
(1938); A. A. Vlasov  Sov. Phys. Usp. \textbf{10}, 721 (1968).

\bibitem{Aleksandrov Rukhadze Book} A. F. Aleksandrov, L. S. Bogdankevich, and A. A. Rukhadze,
\textit{Principles of Plasma Electrodynamics}, Berlin; New York: Springer-Verlag, 1984.


\bibitem{Akhiezer} I. A. Akhiezer, \emph{Plasma
electrodynamics}, (Pergamon Press, 1975).

\bibitem{Landau Vol X} L. Landau and E. M. Lifshitz, \textit{Statistical Physics, Part II}
(Pergamon, New York, 1980).

\bibitem{Zaslavskii JAMTP 66} G. M. Zaslavskii, J. Appl. Mech. Tech. Phys. \textbf{7}, 54 (1966).

\bibitem{Pavlotskii DAN 73} I. P. Pavlotskii, Dokl. Akad. Nauk. USSR \textbf{213}, 812 (1973), (in Russian).

\bibitem{Orlov MM 89} Yu. N. Orlov and I. P. Pavlotsky, Matem. Mod. \textbf{1}, 31 (1989).



\bibitem{Klimontovich book} Yu. L. Klimontovich, \emph{Statistical Physics} Harwood, New York (1986).


\bibitem{Drofa TMP 96}  M. A. Drofa, L. S. Kuz'menkov,
"Continual approach to multiparticle systems with long-range interaction. Hierarchy of macroscopic fields and physical consequences",
Theoretical and Mathematical Physics \textbf{108}, 849 (1996).



\bibitem{Andreev PIERS 2012} L. S. Kuz'menkov and P. A. Andreev,
"Microscopic Classic Hydrodynamic and Methods of Averaging",
presented in PIERS Proceedings, p. 158, Augoust 19-23, Moscow,
Russia 2012.


\bibitem{Kuzmenkov CM 15} L. S. Kuzmenkov, Theoretical Physics: Classical Mechanics (Nauka, Moscow, 2015) [in Russian].

\bibitem{Kuz'menkov 91} L. S. Kuz'menkov,
"Field form of dynamics and statistics of systems of particles with electromagnetic interaction",
Theoretical and Mathematical Physics \textbf{86}, 159 (1991).


\bibitem{Klimontovich Plasma} Yu. L. Klimontovich,
"The Statistical Theory Non-Equilibrium Processes in a Plasma", Pergamon Press, (1967).




\bibitem{Hakim AoP 82} R. Hakim and H. Sivak,
"Covariant Wigner function approach to the relativistic quantum electron gas in a strong magnetic field",
Ann. Phys. \textbf{139}, 230 (1982).


\bibitem{Hakim PRD 92} R. Hakim, L. Mornas, P. Peter, and H. D. Sivak,
"Relaxation time approximation for relativistic dense matter",
Phys. Rev. D \textbf{46}, 4603 (1992).


\bibitem{Shatashvili ASS 97} N. L. Shatashvili, J. I. Javakhishvili, H. Kaya,
"Nonlinear wave dynamics in two-temperature electron-positron-ion plasma",
Astrophys Space Sci. \textbf{250}, 109 (1997).

\bibitem{Shatashvili PoP 99} N. L. Shatashvili and N. N. Rao,
"Localized nonlinear structures of intense electromagnetic waves in two-electrontemperature electron–positron–ion plasmas",
Phys. Plasmas \textbf{6}, 66 (1999).


\bibitem{Hazeltine APJ 2002} R. D. Hazeltine, S. M. Mahajan,
"Fluid description of relativistic, magnetized plasma",
The Astrophysical Journal \textbf{567}, 1262 (2002).


\bibitem{Mahajan PoP 2002} S. M. Mahajan, R. D. Hazeltine,
"Fluid description of a magnetized plasma",
Phys. Plasmas \textbf{9}, 1882 (2002).

\bibitem{Romatschke IJMPE 10} P. Romatschke, "New Developments in Relativistic Viscous Hydrodynamics",
Int. J. Mod. Phys. E, \textbf{19}, 1 (2010).

\bibitem{Mahajan PoP 2011} S. M. Mahajan, Z. Yoshida, "Relativistic generation of vortex and magnetic field",
Phys. Plasmas \textbf{18}, 055701 (2011).


\bibitem{Comisso PRL 14} L. Comisso, F. A. Asenjo, "Thermal-Inertial Effects on Magnetic Reconnection in Relativistic Pair Plasmas",
Phys. Rev. Lett. \textbf{113}, 045001 (2014).



\bibitem{Shatashvili PoP 20} N. L. Shatashvili, S. M. Mahajan , and V. I. Berezhiani,
"Nonlinear coupling of electromagnetic and electron acoustic waves in multi-species degenerate astrophysical plasma",
Phys. Plasmas \textbf{27}, 012903 (2020).



\bibitem{Hakim book Rel Stat Phys} Remi Hakim,
Introduction to Relativistic Statistical Mechanics Classical and Quantum, World Scientific Publishing Co. Pte. Ltd.,  2011.

\bibitem{Asenjo PoP 11} F. A. Asenjo, V. Munoz, J. A. Valdivia, and S. M. Mahajan,
"A hydrodynamical model for relativistic spin quantum plasmas",
Phys. Plasmas \textbf{18}, 012107 (2011).

\bibitem{Melrose BOOK 08} D. B. Melrose, editor. "Quantum Plasmadynamics",
Volume 735 of Lecture Notes in Physics, Berlin Springer Verlag. Springer Verlag, 2008.

\bibitem{Melrose JPA 09} D. B. Melrose and J. I. Weise, "Response of a relativistic quantum magnetized electron gas",
J. Physics A \textbf{42}, H5502 (2009).


\bibitem{Bret PoP 11} A. Bret and F. Haas, "Quantum kinetic theory of the filamentation instability",
Phys. Plasmas \textbf{18}, 072108 (2011).

\bibitem{Ivanov Darwin} A. Yu. Ivanov, P. A. Andreev, L. S. Kuz'menkov,
"Langmuir waves in semi-relativistic spinless quantum plasmas",
Prog. Theor. Exp. Phys. 2015, 063I02 (2015).

\bibitem{Asenjo NJP 12} F. A. Asenjo, J. Zamanian, M. Marklund, G. Brodin, and P. Johansson, "",
New J. Phys. \textbf{14}, 073042 (2012).

\bibitem{Melrose JPA 12} D. B. Melrose, J. I. Weise,
"Spin-dependent relativistic quantum magnetized electron gas",
J. Phys. A \textbf{45}, 5501 (2012).

\bibitem{Ivanov arxiv big 14} A. Yu. Ivanov, P. A. Andreev, L. S. Kuz'menkov,
"Balance equations in semi-relativistic quantum hydrodynamics",
Int. J. Mod. Phys. B \textbf{28}, 1450132 (2014).

\bibitem{Dodin PRA 15 First-principle} D. E. Ruiz, I. Y. Dodin,
"First-principle variational formulation of polarization effects in geometrical optics",
Phys. Rev. A \textbf{92}, 043805 (2015).

\bibitem{Mendonca PoP 11} J. T. Mendonca, "Wave kinetics of relativistic quantum plasmas",
Phys. Plasmas \textbf{18}, 062101 (2011).

\bibitem{Zhu PPCF 12} J. Zhu, P. Ji,
"Dispersion relation and Landau damping of waves in high-energy density plasmas",
Plasma Phys. Contr. Fusion \textbf{54}, 065004 (2012).












\bibitem{Andreev 2021 05} P. A. Andreev, "On the structure of relativistic hydrodynamics for hot plasmas", arXiv:2105.10999.




\bibitem{Andreev 2021 09} P. A. Andreev,
"Microscopic model for relativistic hydrodynamics of ideal plasmas",
arXiv:2109.14050.

\bibitem{Andreev 2021 10} P. A. Andreev,
"Anisotropic pressure effects in hydrodynamic description of waves propagating parallel to the magnetic field in relativistically hot plasmas",
arXiv:2110.14749.


\bibitem{Andreev 2112} P. A. Andreev,
"Spin-electron-acoustic waves and solitons in high-density degenerate relativistic plasmas",
arXiv:2112.13880.

\bibitem{Andreev 2202} P. A. Andreev,
"Nonlinear Coupling of Electromagnetic and Spin-Electron-Acoustic Waves in Spin-polarized Degenerate Relativistic Astrophysical Plasma",
arXiv:2202.11814.





















\end{thebibliography}
\end{document}